\documentclass[acmtog]{acmart}
\usepackage{booktabs}
\usepackage[ruled]{algorithm2e}

\SetAlFnt{\small}
\SetAlCapFnt{\small}
\SetAlCapNameFnt{\small}
\SetAlCapHSkip{0pt}
\usepackage{multirow}
\usepackage{makecell}
\usepackage{algpseudocode}
\usepackage{xcolor}

\acmJournal{TOG}

\begin{document}

\setcopyright{acmlicensed}
\acmJournal{TOG}
\acmYear{2025} \acmVolume{44} \acmNumber{6} \acmArticle{} \acmMonth{12}\acmDOI{10.1145/3763311}

\title{Shape-aware Inertial Poser: Motion Tracking for Humans with Diverse Shapes Using Sparse Inertial Sensors}

\author{Lu Yin}
\affiliation{%
 \institution{Xiamen University}
 \department{School of Informatics}
 \city{Xiamen}
 \country{China}}
 \email{yinlu@stu.xmu.edu.cn}
 
\author{Ziying Shi}
\affiliation{%
 \institution{Xiamen University}
 \department{School of Informatics}
 \city{Xiamen}
 \country{China}}
 \email{23020241154433@stu.xmu.edu.cn}

\author{Yinghao Wu}
\affiliation{%
 \institution{Xiamen University}
 \department{School of Informatics}
 \city{Xiamen}
 \country{China}}
 \email{30920231154365@stu.xmu.edu.cn}

\author{Xinyu Yi}
\affiliation{%
 \institution{Tsinghua University}
 \department{School of Software and BNRist}
 \city{Beijing}
 \country{China}}
 \email{yixy20@mails.tsinghua.edu.cn}

\author{Feng Xu}
\affiliation{%
 \institution{Tsinghua University}
 \department{School of Software and BNRist}
 \city{Beijing}
 \country{China}}
 \email{feng-xu@tsinghua.edu.cn}

\author{Shihui Guo*}
\affiliation{%
 \institution{Xiamen University}
 \city{Xiamen}
 \country{China}
}
\email{guoshihui@xmu.edu.cn}

\renewcommand\shortauthors{Lu Yin, Ziying Shi, Yinghao Wu. et al}

\begin{abstract}

Human motion capture with sparse inertial sensors has gained significant attention recently. However, existing methods almost exclusively rely on a template adult body shape to model the training data, which poses challenges when generalizing to individuals with largely different body shapes (such as a child). This is primarily due to the variation in IMU-measured acceleration caused by changes in body shape. To fill this gap, we propose Shape-aware Inertial Poser (SAIP), the first solution considering body shape differences in sparse inertial-based motion capture. Specifically, we decompose the sensor measurements related to shape and pose in order to effectively model their joint correlations. Firstly, we train a regression model to transfer the IMU-measured accelerations of a real body to match the template adult body model, compensating for the shape-related sensor measurements. Then, we can easily follow the state-of-the-art methods to estimate the full body motions of the template-shaped body. Finally, we utilize a second regression model to map the joint velocities back to the real body, combined with a shape-aware physical optimization strategy to calculate global motions on the subject. Furthermore, our method relies on body shape awareness, introducing the first inertial shape estimation scheme. This is accomplished by modeling the shape-conditioned IMU-pose correlation using an MLP-based network. To validate the effectiveness of SAIP, we also present the first IMU motion capture dataset containing individuals of different body sizes. This dataset features 10 children and 10 adults, with heights ranging from 110 cm to 190 cm, and a total of 400 minutes of paired IMU-Motion samples. Extensive experimental results demonstrate that SAIP can effectively handle motion capture tasks for diverse body shapes. The code and dataset are available at \href{https://}{\textit{https://github.com/yinlu5942/SAIP}}.

\end{abstract}
\maketitle
\vspace{-0.3\baselineskip}

\begin{center}
\begin{minipage}{0.5\textwidth} 
  \centering
  \includegraphics[width=\linewidth]{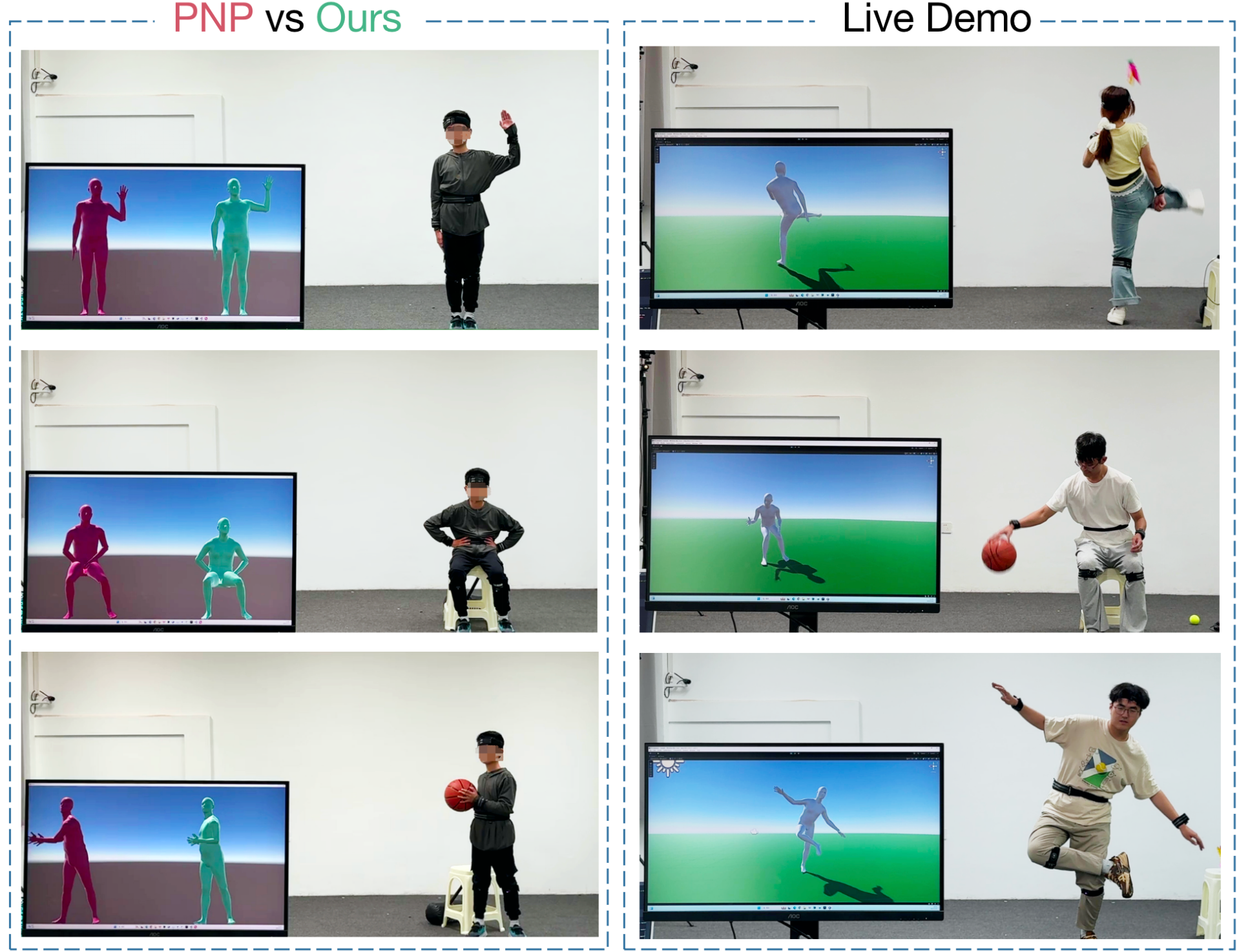}
  \captionof{figure}{ \textbf{Left}: Live comparison on a child subject between the state-of-the-art inertial motion capture system PNP \cite{yi2024pnp} (red) and our method (green). Our solution effectively handles the child character, whereas PNP shows errors. \textbf{Right}: Additional live demonstrations of our method showcase its capability to handle complex motions across diverse body shapes.}
  \Description{}
  \label{fig:teaser}
\end{minipage}
\end{center}

\section{Introduction}Human motion capture is a highly promising technique that is showing increasing impact across fields such as VR/AR, embodied intelligence, rehabilitation, and animation. Motion capture with dense markers or wearable sensors~\cite{point2011optitrack, noitom2017perception, paulich2018xsens} has demonstrated very high precision but requires heavy systems that are not applicable to end users. Image-based solutions~\cite{zhang2021direct, yu2021function4d, sun2019deep, ye2022faster, zhao2024single} are much more lightweight but suffer from occlusions and challenging lighting, and they are not suitable for everyday use, as camera shooting is always required. Recently, there has been a new trend to use sparse Inertial Measurement Units (IMUs) for motion capture ~\cite{von2017sparse, huang2018deep, yi2021transpose, yi2022physical, jiang2022transformer, asip, yi2024pnp}. Using only six IMUs placed on key body parts (hands, legs, head, and waist), they significantly enhance comfort and portability, offering great potential for everyday motion capture tasks.

However, due to the scarcity of real-world data, state-of-the-art sparse IMU-based systems are predominantly trained on synthesized IMU data derived from a \textit{template human body model}. Consequently, existing methods implicitly approach inertial motion capture by only modeling the relationship between \textit{human pose and IMU signals}. We argue that this is inadequate, as IMU signals are influenced not only by human motion but also by shape variations. While these systems perform well on many adult characters,
their accuracy significantly declines when applied to subjects like children, whose body shapes substantially deviate from the training samples (see Fig.~\ref{fig:teaser}).

The identified issue primarily arises from shape-conditioned IMU measurements, where positional kinematic data (e.g., position, velocity, acceleration of joint and mesh) varies due to changes in body shape. The impact of body shape on IMU signals manifests in various ways (see Fig.~\ref{fig:shape_acc}): for instance, acceleration during rotation around a parent joint is primarily influenced by bone length (rotation radius), while rotation around the bone itself is affected by limb fatness and other shape attributes. In contrast, other motions such as jumping or falling, are largely unaffected by shape variations in positional data, leading to specific scenarios. The complexity of shape-conditioned IMU measurements underscores the importance of modeling the relationship between \textit{body shape and IMU signals}, a critical task in inertial motion capture. To model this relationship, one potential solution is to retrain the system using synthesized IMU-motion data across diverse body shapes, though this approach complicates training due to the need to learn the triadic relationship among body shapes, poses, and IMU signals. Alternatively, we propose disentangling this triadic relationship, offering a more efficient way to address the challenge.

In this paper, we address errors in inertial motion capture caused by body shape variations by proposing the first shape-aware solution, which independently models the effects of body shapes and poses on IMU signals. Our approach involves two key steps: \textit{1)} modeling the correlation between IMU signals and body shape by mapping IMU measurements from diverse real body shapes to a template body shape under identical poses, and \textit{2)} leveraging an established pose estimation framework designed for the template body shape to model the relationship between IMU signals and motion. To accomplish this, we first propose a learning-based kinematic signal retargeting method for step \textit{1)}. Specifically, we synthesize IMU measurements across various body shapes under identical poses, and train a neural network to map the input IMU data from a real human body to the corresponding IMU data of a template adult body model. In step \textit{2)}, we adopt state-of-the-art techniques to decompose the motion capture task into local pose estimation and global movement estimation. While the local pose regressed from the template body's IMU data aligns with the real body shape, the global movement requires recalculation. To address this, we utilize a second retargeting network to regress the real body's joint velocities from those of the template body and implement a shape-aware physics optimization module to calculate the global movements of the real body. Furthermore, the retargeting nets require awareness of the real body shape. To address this, we introduce the first human shape estimation scheme in a sparse IMU-only system by modeling shape-conditioned IMU signals. We develop an MLP-based estimation algorithm that uses IMU data, pose, and body height as inputs. Since our motion tracking framework relies on shape input, we initialize the shape using body height for the first window's pose estimation, then iteratively refine the predicted shape and pose over time.

Additionally, to validate our SAIP method and enrich open-source IMU-based motion datasets, we present a multi-shape inertial motion capture dataset. This dataset comprises recordings from 10 adults over 18 years old and 10 children aged 5-10, totaling 1.5 million frames (over 7 hours) of diverse activities, including sports, daily actions, and freestyle movements. The dataset features subjects ranging from young children as short as 118 cm to adults exceeding 190 cm, covering a broad spectrum of body sizes. 

In summary, our contributions are:

\begin{itemize}
\item The first shape-aware sparse inertial human motion capture solution, achieving real-time motion tracking for diverse shapes (including small children), which we call SAIP.
\item A learning-based approach to bidirectionally regress positional signals (acceleration and velocity) between various-shaped bodies and a template SMPL body model, addressing the kinematic signal difference affected by human shapes.  
\item Shape acquisition is achieved through modeling the shape-pose conditioned IMU signals using an MLP-based algorithm. We are the first system to estimate human shape using only sparse IMUs and body height input. 
\item  A Multi-shaped Inertial Motion Capture Dataset (MID) with IMU and ground truth motion data collected from 20 various-shaped subjects, which is also the first IMU database with examples of pre-teen children.  
\end{itemize}

\section{Related Work}

\subsection{Human Motion Tracking using IMU sensors}
  Motion reconstruction using IMUs typically involves attaching the IMUs to key body parts and solving inverse kinematics (IK) based on IMU measurements to obtain joint rotations (body pose). In commercial systems such as Xsens~\cite{paulich2018xsens} and Noitom PN Series~\cite{noitom2017perception},  human pose estimation using dense IMUs (e.g. 17 IMUs) have achieved high accuracy. 
  
  In recent years, methods that use sparse IMUs (e.g., only 6 IMUs) attached to the arms, legs, head, and waist have garnered significant attention. Huang~\shortcite{huang2018deep} was the first to use recurrent neural networks (RNNs) to estimate human pose in an end-to-end manner. Yi~\shortcite{yi2021transpose} introduced a multi-stage prediction framework that incorporates joint position information to estimate more accurate human pose, achieving global motion tracking as well. In subsequent studies, Jiang~\shortcite{jiang2022transformer} and Wu~\shortcite{asip} employed the transformer architecture to improve pose estimation accuracy, while Yi~\shortcite{yi2022physical} proposed a physics-based optimization method to calculate global movement and make the predicted human motion physically plausible. Most recently, Yi~\shortcite{yi2024pnp} addressed the challenge of modeling non-inertial forces when the root joint operates in a non-inertial coordinate system, thereby correcting acceleration measurements to achieve more accurate motion tracking under acceleration-domain motions.
  
  Some studies also apply IMU data using nonattached techniques. For example, Zuo~\shortcite{Zuo_2024_CVPR} estimate upper body pose using IMUs embedded in a loose-wear jacket for a comfort user experience.  Another widely followed approach involves integrating additional 3D position information alongside IMU data. Using VR device trackers with cameras or external stations such as HMDs (Head Mounted Devices) and controllers, some studies can generate full-body motions from 3 trackers on head and wrists ~\cite{Jiang2022AvatarPoserAF, du2023agrol, lee2023questenvsim, lee2024mocap, liang2023hybridcap, dittadi2021full, jiang2024egoposer, aliakbarian2022flag, winkler2022questsim, yang2024divatrack, yang2021lobstr, feng2024stratified}, while Ponton~\shortcite{ponton2023sparseposer} uses 6 trackers to perform high accuracy motion tracking.
  However, most existing methods rely on a template SMPL \cite{loper2023smpl} body model to represent output motion. While this approach performs well for many adult subjects, it struggles to accommodate a diverse range of body shapes, particularly those of small children.

\subsection{Human Shape Estimation}

Human shape estimation involves reconstructing human meshes from real-world data inputs. Pioneering efforts in this field relies on optical data, such as images and videos, which provide rich information about body shape \cite{yang2021viser, kocabas2020vibe, chibane2020implicit, li20213d, li2021hybrik, pang2022benchmarking, sengupta2020synthetic, shen2023global, cai2023smpler}. However, other applications—including body-worn sensor-based pose estimation, human action recognition, and humanoid control—also require shape information, yet lack access to camera-based inputs. Recent studies have explored alternatives for shape estimation in sensor-based systems; for instance, Yang~\shortcite{yang2024divatrack} propose a calibration process to adapt human skeletons, while Jiang~\shortcite{jiang2024egoposer} leverage head-mounted displays (HMDs) to estimate shape proportions. Nevertheless, these approaches rely on additional data, such as controller positions or HMD-derived images, restricting their applicability to specific system configurations.

\subsection{Motion Capture Datasets}
  Human motion data typically consists of a sequence of poses that describe a person’s actions, while many applications also emphasize global movement expressed in 3D coordinates. With the rapid progress in deep learning, extracting human poses from videos and RGB images to derive motion data has become a dominant approach. Many large-scale motion datasets also rely on dense marker systems \cite{plappert2016kit, lin2023motion, kratzer2020mogaze, chatzitofis2020human4d, black2023bedlam, AMASS}. In addition to optical systems, inertial sensors offer another effective method for acquiring high-precision human motion data. In recent years, the research community has seen the emergence of several open-source IMU databases, such as \cite{huang2018deep, trumble2017total, maurice2019human, palermo2022complete}. 
  Although various motion capture datasets encompass a broad spectrum of body shapes, most lack child subjects. Datasets that do include children, such as \cite{aloba2018kinder}, suffer from noise and suboptimal quality. The dataset proposed in \cite{dong2020adult2child} collects motion data from 8 children for a style transfer task, but it is limited to basic actions like walking, running, and jumping, lacking motion diversity.

\section{Method}

Our task is to reconstruct human motion from 6 IMUs attached to the human body. The input is each IMU's acceleration ${A}\in \mathbb{R}^{3}$, angular velocity ${\omega}\in \mathbb{R}^{3} $, and orientation $ R \in \mathrm{SO}(3)$. The outputs are the human pose defined as joint rotations using the 6D~\cite{zhou2019continuity} representation $\theta \in \mathbb{R}^{6n}$ and global translation of the root joint ${r}_{root}\in \mathbb{R}^{3}$, where $n = 24$ denotes the number of the body joints in the SMPL~\cite{loper2023smpl} skeleton. 

\subsection{Shape-conditioned IMU Signals}

Previous work often assumes a mean body shape when estimating motion from sparse IMUs. However, we argue that body shape significantly influences IMU measurements, where large inconsistencies between the assumed and actual body shape can substantially degrade motion estimation accuracy. For instance, when capturing a child's movement, IMUs generate smaller acceleration compared to an adult's. Interpreting these measurements based on an adult's body shape will produce incorrect motion dynamics.

\begin{figure}[t]
	\centering
 \includegraphics[width=0.9\linewidth]{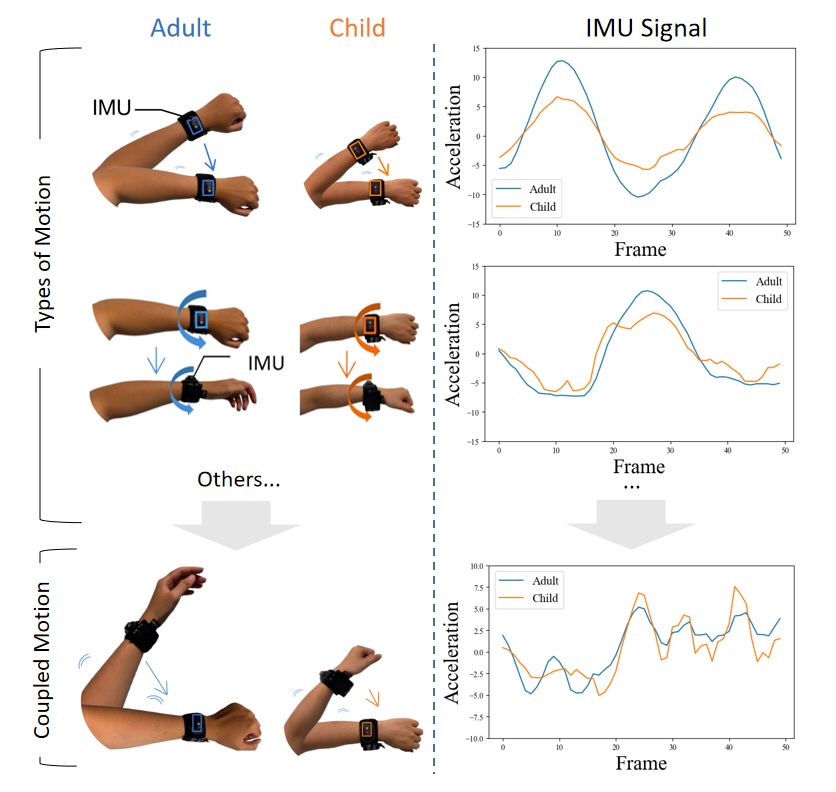}
	\caption{ Illustration of Our Motivation: IMU signals in real-life actions are influenced by various shape-related factors. Without kinematic signal retargeting, current baseline methods, trained solely on adult data, fail to accurately perform inertial motion tracking for subjects with diverse shapes, as the input signals deviate significantly from the training samples.
}
	\label{fig:shape_acc}
\end{figure}

Modeling the influence of body shape on IMU signals is nontrivial. As illustrated in Fig.~\ref{fig:shape_acc}, when individuals with different body shapes perform the same motion (i.e., identical joint rotations), their acceleration patterns can vary significantly and cannot be approximated by a simple scale factor. For instance: motion like 1) body swing induces accelerations proportional to local bone length, and 2) body twist produces accelerations correlated with body fatness; while other motion like 3) jumping can yield shape-invariant accelerations (e.g., gravitational acceleration). Thus, real-world IMU signals couple multiple aspects of body shape, e.g. bone lengths and fat distribution, in a pose-dependent manner, making their interpretation highly context-dependent.

\begin{figure*}[t]
	\centering
 \includegraphics[width=\linewidth]{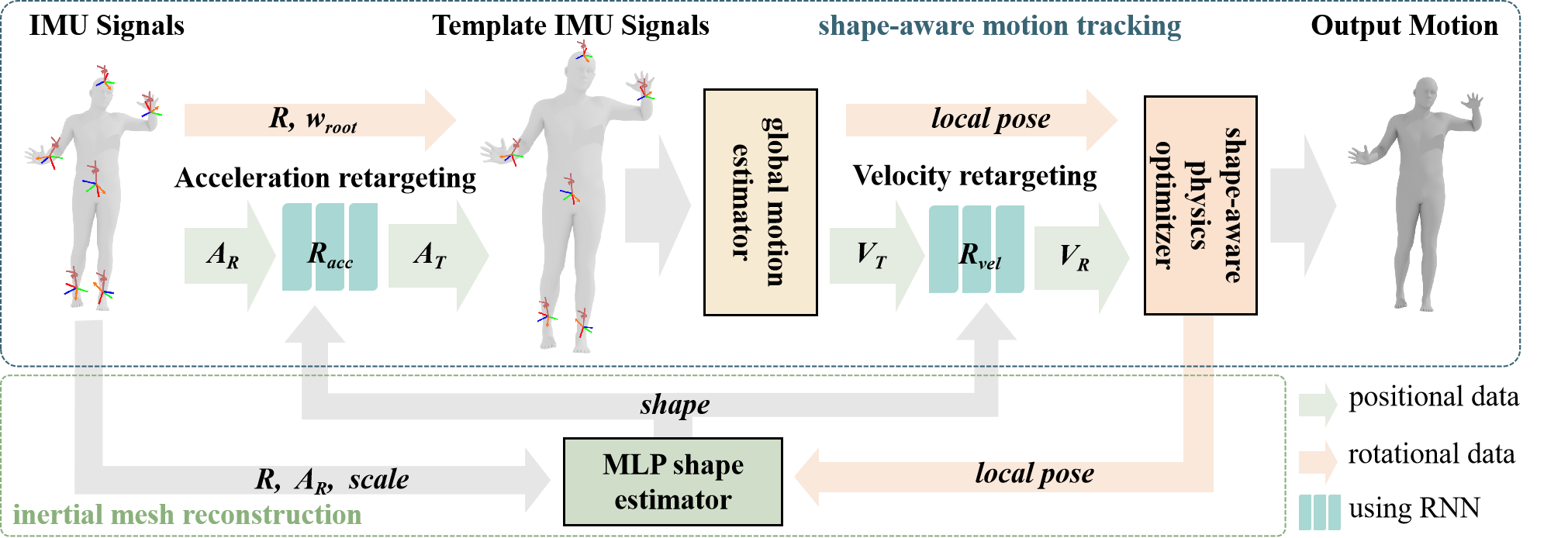}
	\caption{Pipeline of Our Method: 
\textit{1) Shape-Aware Motion Tracking}: Real IMU signals are first mapped to a template-shaped model using $R_{acc}$. Next, a global motion estimator regresses local pose and joint velocities, followed by $R_{vel}$ to map the estimated velocities back to the target character. Our shape-aware motion optimizer employs a dynamic model tailored to the real body shape to derive physically accurate global motion.  
\textit{2) Inertial Mesh Reconstruction}: We utilize an MLP-based shape estimator to achieve shape awareness by leveraging IMU signals and the estimated pose.}
	\label{fig:overview}
\end{figure*}

\subsection{Framework Overview}

Our method consists of two core components:

\textbf{Shape-Aware Motion Tracking} (Fig.~\ref{fig:overview}, top half) captures shape-dependent human motion from IMU signals and body shape information. To decouple shape and motion in the IMU measurements, we first retarget the IMU signals to a template model with a mean body shape while preserving the motion (Kinematic Signal Retargeting Module, Section 3.2). These shape-invariant signals are then processed by an off-the-shelf motion estimator, yielding pose and translation estimates aligned with the mean shape. Next, we fix the estimated pose and retarget the translation back to the real shape using a velocity-based retargeting network. Finally, to enhance physical plausibility, we apply a shape-aware physics-based optimization to refine the reconstructed motion (Section 3.3).

\textbf{Inertial Mesh Reconstruction} (Fig.~\ref{fig:overview}, bottom half) 
progressively refines the estimated body shape using both the reconstructed motion and input IMU measurements (Section 3.4).

The two components operate synergistically: Shape-Aware Motion Tracking provides accurate kinematic data for mesh refinement, while Inertial Mesh Reconstruction improves shape estimation which in turn enhances motion tracking accuracy. Additionally, we introduce the \textbf{MID Dataset}, a novel collection of IMU and motion data under diverse body shapes to facilitate evaluation (Section 3.5).

\subsection{Learning-based Kinematic Signal Retargeting}

Baseline methods for motion capture typically decompose the task into two subtasks: local pose estimation and global movement estimation. These methods take IMU measurements as input and output local poses and global joint velocities. To address the challenges outlined in Section 3.1.1, we propose the following approach:

The first kinematic signal retargeting network, denoted as $R_{acc}$ in Fig.~\ref{fig:overview}, takes positional IMU data (acceleration $A_R$) and body shape information $\beta$ as inputs and regresses the acceleration $A_T$ corresponding to the template body shape. We adopt the pose estimation component from state-of-the-art PNP \cite{yi2024pnp}, using the standardized IMU data as input to infer local poses and joint velocities $V_T$. Here, local poses are represented by joint rotations, which are not affected by body shape variations.
Analogous to $R_{acc}$, a second kinematic signal retargeting network, $R_{vel}$, takes body shape information as input and maps the joint velocities to the target body shape. Through this process, we obtain the pose and global joint velocities of the target body. To model the temporal nature of IMU signals, $R_{acc}$ and $R_{vel}$ employ recurrent neural networks (see our supplementary paper for implementation details).

The inputs and outputs of the retargeting networks consist of shape-conditioned positional data (acceleration and velocity). Such ground truth data is evidently unavailable in real-world scenarios. 
To train $R_{acc}$ and $R_{vel}$, we need:

\begin{enumerate}
    \item A set of diverse body shapes, including not only SMPL shapes but child characters. 
    \item Motion data corresponding to these body shapes, as different global movements occur for individuals performing the same action.
    \item Simulated joint acceleration and velocity data derived from motion data.
\end{enumerate}

AMASS \cite{AMASS} dataset contains motion data for adults and their corresponding SMPL body shapes. To enrich body shapes, we scale the original  shapes obtained from AMASS within a range of 0.5 to 1.2, resulting in shapes with heights ranging from 0.8 to 2.0 m, covering a much wilder range of human bodies, including children subjects. 
To compute the correct global movement under body shape variations, we adjust the global velocity of the root joint based on foot-ground contact. For a given motion sequence and its paired body model, we calculate forward kinematics (FK) to obtain four mesh vertices' positions ${v}_i$ at the tips and heels of both feet in frame $i$.  The initial position $\text{tran}_0^T$ of the target body $T$ is aligned with the SMPL body $M$: the $x$- and $z$-coordinates of $\text{tran}_0^T$ match those of $\text{tran}_0^M$, while the $y$-coordinate (vertical position) is also determined by the scale, ensuring foot-ground contact. 
For the rest of the motion sequence, if $\|{v}_i - {v}_{i-1}\| < 0.5 \, \text{cm}$, we consider the vertex to be stationary (i.e., in contact with the ground). For each frame $i$, if any vertex satisfies the stationary condition, we calculate the velocity of that vertex $V_i$ (in the root joint frame). Since the vertex is stationary, translation of the root joint is updated as: 
\begin{equation}
    \text{tran}_i^T = \text{tran}_{i-1}^T + {V}_i.
\end{equation}
When neither foot satisfies the stationary condition (e.g., during a jump), the root joint global position is updated as:
\begin{equation}
    \text{tran}_i^T = \text{tran}_{i-1}^T + {V}_i^M,
\end{equation}
where the scaled body and the SMPL body have identical root joint velocities during jumping motions. This ensures that the IMU acceleration measurements correspond to gravitational acceleration. Finally, we use the acceleration synthesis algorithm based on energy optimization proposed in ~\cite{yi2024pnp} to generate the required IMU acceleration data $A_R$ and $A_T$.

\subsection{Global motion reconstruction}
We conduct global motion estimation using the regressed acceleration $A_T$, alongside other IMU measurements such as orientation and angular velocity. Initially, we adopt the method from \cite{yi2024pnp} to model non-inertial acceleration, employing a neural auto-regressive estimator to learn the physically accurate fictitious forces resulting from the non-inertial root coordinate frame of the human body, yielding the fictitious force acceleration. Next, following the framework in \cite{yi2021transpose}, we utilize three neural networks to sequentially predict: the positions of five leaf-node joints, full-body joint positions, and joint rotations. For global movement estimation, we apply the approach from \cite{yi2022physical} to regress joint velocities $V \in \mathbb{R}^{24 \times 3}$ and foot-ground contact probabilities.

Our implementation of a learning-based kinematic signal retargeting method and global motion estimator (Section 3.1) effectively estimates local pose and global movement by independently modeling the influence of body shape variations and IMU signals on the output motion. Subsequently, state-of-the-art approaches~\cite{yi2022physical} employ a physics-based optimization strategy, leveraging the estimated local pose, foot-ground contact, and joint velocities to produce more physically plausible human motion. However, this strategy relies on a fixed adult body model~\cite{shimada2020physcap}, limiting its adaptability to diverse body shapes. To overcome this, we introduce a dynamic model that integrates body shape information. For a human body with known shape parameters $\beta$, we compute physical properties such as mass, center of mass, and inertia, enabling the original optimization strategy to adaptively control characters with varying body shapes. We will elaborate on this physics-based optimization technique in Section.~\ref{sec:phys}.

\begin{algorithm}
\caption{Inertial Mesh Reconstruction}
\label{alg:refine}
\KwData{IMU acceleration $A_R$, orientation $R$, subject body height $H_R$, template body height $H_T$, template body shape $\beta_0$, refine time window $W$.}
\KwResult{Sequence of local poses $\theta = \{\theta_1, \theta_2, \dots\}$, SMPL shape parameters $\beta$}
Initialize $\text{scale} \gets H_R / H_T$, $\beta \gets \beta_0 * \text{scale}$, $\theta \gets []$, $t \gets 1$, $W \gets 60$;
\While{$A_R$ and $R$ is available}{
    $\theta_t \gets \text{PoseEstimator}(A_R, \beta, \dots)$
    // \textcolor{gray}{Estimate pose for current frame.}\\
    $\theta \gets \theta \cup \{\theta_t\}$
    // \textcolor{gray}{Append pose to sequence.}\\
    \If{$t \mod W = 0$}{
    $\beta \gets \text{ShapeEstimator}(A_R, \text{scale}, \theta^{t-W+1\rightarrow t})$
    // \textcolor{gray}{Update shape parameters every window}\\
    }
}
\end{algorithm}

\subsection{Inertial Mesh Reconstruction}
In Section 3.1, we analyzed how body shape variations affect acceleration under identical poses, leveraging body shape and the target character’s acceleration to regress local pose. Likewise, shape-conditioned acceleration variations encode body shape information, which we exploit using an MLP to regress SMPL shape parameters $\beta$ from $A_R$ and the predicted pose. We employ a 60-frame (1-second) window to update body shape. Our system uses the subject’s body height to compute a scale factor relative to the template body height. This scaled zero shape initializes the retargeting networks in the first window, producing a size-aware-only initial local pose. Subsequently, our MLP estimates an initial body shape from the pose and IMU signals within this window. Using these estimated shape parameters, the motion tracking algorithm performs shape-aware motion estimation in the second window. As the number of windows increases, both body shape and pose estimation improve in accuracy (Alg. \ref{alg:refine}).

\subsection{Shape-aware Dynamic Model} 
\label{sec:phys}
Physical optimization is employed to derive globally consistent motion that adheres to physical constraints based on kinematic estimations. While some studies utilize optimization-based techniques \cite{andrews2016real, rempe2020contact, li2019estimating, shimada2020physcap} to determine optimal forces and human motion that comply with physical laws, such as the equation of motion \cite{featherstone2014rigid}, others leverage reinforcement learning \cite{schreiner2024adapt, bergamin2019drecon, isogawa2020optical, yu2021human, yuan2019ego, yuan2021simpoe} in physics-based character control, harnessing advanced non-differentiable physics simulators. Among these, our approach most closely aligns with that of Yi et al.~\cite{yi2022physical}, which implements a dual PD (Proportional-Derivative) controller for joint rotations and positions. This method introduces a novel dual PD controller to enhance global character control and accuracy, marking the first explicit integration of physics-based optimization into sparse IMU-based motion capture. However, discrepancies between joint velocities and the fixed physical properties of the dynamic model hinder precise global motion control. To address this, we propose a shape-aware enhancement to this approach.

The joint positions and the mesh vertex positions of the human body in the initial state for individuals with different body shapes are first obtained by calculating the forward kinematics (FK). Then, the body is sliced along the x-axis (in SMPL coordinates, x-left, y-up, z-formward) with a defualt step size $res$ (resolution for mesh voxelization) of 2 cm, where the global inner points $P$ are determined.  The weight of points is calculated using linear interpolation. Whereas the weighted mass of the points, as seen from joints is: 
\begin{equation}
w_{ij} = \text{weight}_{ij} * {res}^3
\end{equation}
Then, mass of joint $i$ is:
\begin{equation}
    m_i = \sum_{j=1}^M w_{ij},     
\end{equation}
where $M$ is total number of inner points and $w_{ij}$ is mass weight of joint $i$ as seen from point $j$.  We then compute centor of mass (com) of joint $i$ as:
\begin{equation}
    c_i = \frac{\sum_{j=1}^M w_{ij} (P_i - j_i )}{m_i} 
\end{equation}
Next, we have the contribution of all points to the center of mass of joint $i$ defined as the inertia:
\begin{equation}
    I_i = \sum_{j=1}^M w_{ij}R_jR_j^T, 
\end{equation}
where $R_j$ is the offset vector of point $j$ relative to the com, denoted as $[P_j - c_i]_{\times}$
We incorporate body shape-informed physical properties mass, com and inertia into the dual PD controller to achieve our shape-aware physical optimization strategy. Physical optimization is implemented using the Rigid Body Dynamics Library (RBDL)\cite{Felis2016}.

\begin{figure}[t]
	\centering
 \includegraphics[width=1\linewidth]{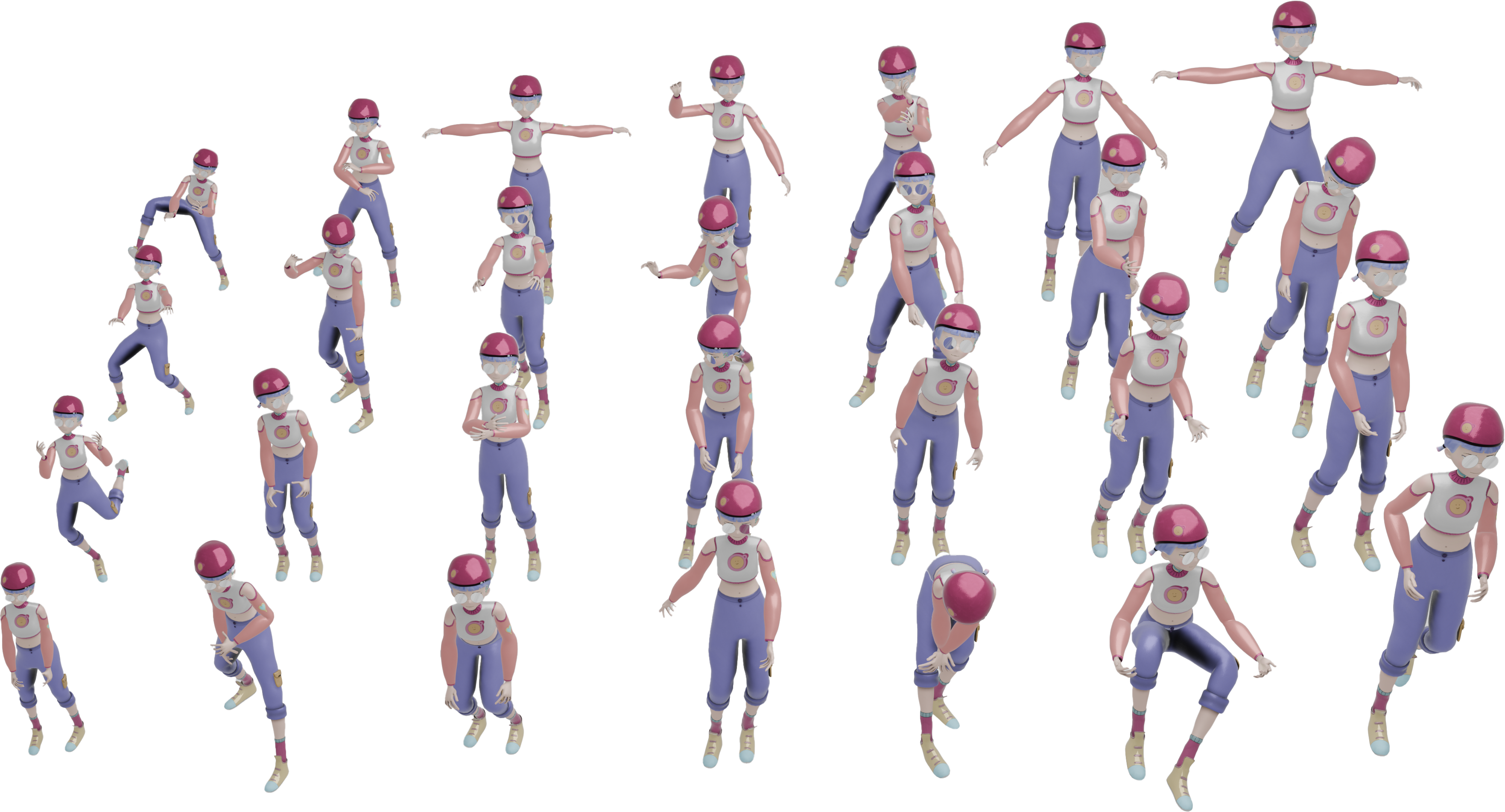}
	\caption{Our MID dataset contains IMU-Motion paired samples collected from 20 subjects with diverse shapes.}
	\label{fig:dataset}
\end{figure}

\subsection{Multi-shape Inertial MoCap Dataset (MID)}

We introduce the Multi-shape Inertial MoCap Dataset (MID), which, to our knowledge, is the first IMU dataset in the research community to include pre-teen children. The dataset comprises 20 participants: 10 children aged 5–10 years and 10 adults over 18 years, with heights ranging from 110 cm to over 190 cm, ensuring a diverse range of body shapes (Fig.~\ref{fig:dataset}). 
\subsubsection{Consent}
All participants signed an informed consent form before joining the study. For child participants, we provided an age-appropriate consent version, while their parents received the complete consent form before motion capture commenced. Additionally, all live demonstration recordings and the use of images in our supplementary video were conducted with explicit written consent from all participants.
\subsubsection{Motion Capture}
Data collection was performed using Noitom’s PN series sensors and the PN Studio system ~\cite{noitom2017perception}, which utilizes 17 PN IMU sensors attached at fixed positions to measure IMU data and compute joint rotations. Prior to data collection, each participant’s bone lengths were measured and used by the PN Studio software to construct skeletal models, enabling accurate pose and global movement capture. 
We opted for a 17-IMU motion capture system over marker-based systems due to two key limitations of the latter: 1) marker-based systems require custom-tailored MoCap suits for children, as commercial options are designed solely for adults, and 2) the active nature of children often leads to markers detaching during capture, complicating data processing. In contrast, the 17-IMU system adapts to diverse body shapes without requiring a suit, as IMUs are simply attached to fixed positions, ensuring stability and ease of use. In comparison to multiview camera systems, the 17-IMU system provides more accuracy and represents the most precise reference we could obtain, as chosen by other datasets in the research community (such as Nymeria \cite{ma2024nymeria} and DIP-IMU \cite{huang2018deep}).
During data collection, we first explained the process to child participants and demonstrated MoCap results to engage their interest and attention. Given children’s typically lower compliance, we avoided prescribing specific actions, instead encouraging freestyle movements. This approach yielded a diverse, child-styled motion dataset, reflecting varied expressions across different body shapes (Fig.~\ref{fig:dataset}). Moreover, the 17-IMU system’s independence from camera setups and optical constraints enabled data collection in unconstrained outdoor environments, such as playgrounds and basketball courts. Consequently, our dataset includes extended global movement sequences, featuring minutes-long recordings of children walking and running.

\subsubsection{Dataset Composition}
The MID dataset provides the following:  
1. \textit{Motion Data}: Raw motion data files are exported from PN Studio in BVH and FBX formats, based on Noitom’s default skeletal structure. Additionally, we provide processed motion data aligned with the SMPL skeleton, including local pose and global root joint position data.  
2. \textit{IMU Data}: Raw IMU data, recorded in the sensor coordinate system, are calibrated to the SMPL coordinate system for inertial motion capture applications. The dataset includes acceleration, angular velocity, and orientation data from 17 IMUs ( featuring the 6 IMUs used in our sparse inertial motion capture task) in CSV format at 60 FPS.  
In total, the MID dataset contains 1.5 million frames (over 7 hours) of motion and IMU data. We also leverage this dataset to validate our proposed SAIP method.

\section{Experiments}

\subsection{Implementation Details}

\begin{table}[ht]
    \centering
    \caption{Quantitative comparison results with state-of-the-art. The transofmer-based method ASIP~\cite{asip} is retrained using our augmented data containing subjects with diverse shapes. }
    \label{tab:comparison}
    \begin{minipage}{0.48\textwidth}
        \centering
        \begin{tabular}{c|ccccc}
            \toprule
            Method & SIP Err & Ang Err & Joint Err & Mesh Err & Jitter \\
            \midrule
            \multicolumn{6}{c}{DanceDB*} \\
            \midrule
            TransPose & 47.06& 32.83& 19.85& 23.31& 4.17\\
            PIP & 19.48& 12.16& 8.20& 9.58& \textbf{0.32}\\
            TIP & 20.14& 13.59& 8.48& 9.70& 0.80\\
            ASIP & 17.68& 12.10& 7.98& 8.46& 0.71\\
            PNP & 15.58& 10.46& 6.90& 8.02& 0.86\\
            SAIP (ours) & \textbf{12.23}& \textbf{8.27}& \textbf{5.17}& \textbf{5.96}& 0.35\\
            \midrule
            \multicolumn{6}{c}{MID} \\
            \midrule
            TransPose & 26.29& 15.68& 8.40& 9.08& 0.20\\
            PIP & 25.10& 13.59& 9.54& 11.52& \textbf{0.09}\\
            TIP & 31.68& 17.99& 9.76& 11.14& 0.27\\
            ASIP & 22.20& 13.75& 8.81& 9.42& 0.16\\
            PNP & 23.22& 14.60& 8.18& 9.09& 0.10\\
            SAIP (ours) & \textbf{21.00}& \textbf{8.67} & \textbf{5.24} & \textbf{6.09} & 0.12\\
            \bottomrule
        \end{tabular}
    \end{minipage}%
    \hfill
\end{table}

\textbf{Datasets}  We utilize the augmented AMASS dataset (Section 3.2) to synthesize IMU data for training. The augmented DanceDB dataset \cite{AMASS_DanceDB} (which we call DanceDB*), featuring child body shapes with heights ranging from 0.8 m to 1.2 m, serves as one of the test sets. The shape estimator is also trained on the AMASS dataset. Additionally, we validate our method on our real-world MID dataset and the TotalCature\cite{trumble2017total} dataset. 

\noindent \textbf{Metrics} We use the same metrics as in \cite{yi2021transpose, asip, yi2022physical, yi2024pnp}
to evaluate motion tracking accuracy: \textit{1) SIP Errors (degrees)} measures mean global orientation error of shoulders and hips, \textit{2) Angular Error (degrees)} measures mean global rotation error for all body joints, \textit{3) Positional Error (cm)} measures mean position error of all body joints, \textit{4)} \textit{Mesh Error (cm)} measures mean vertex position error of all SMPL meshes, and \textit{5) Jitter Error ($10^3 $m/s$^3$)} measures mean body joints jerk.  We alse evaluate body shape esimating results using \textit{6) Mesh Error-T} (cm): mesh error in the T-pose. For all of the above metrics, smaller values indicate higher accuracy. 

\noindent \textbf{Network Setup} The retargeting networks $R_{acc}$ and $R_{vel}$, are implemented as Recurrent Neural Networks (RNNs), comprising a linear input layer, two long short-term memory (LSTM) layers, and a linear output layer. The linear layers employ ReLU as the activation function, with the hidden layer dimension set to 256. For both RNNs, a dropout rate of 40\% is applied and the batch size is set to 256. The shape estimator contains 4 layers with a hidden width of 512. All networks are optimized using the ADAM optimizer in training. In the template-shaped global motion estimation stage, we follow \cite{yi2024pnp}, utilizing 5 RNNs to regress local poses and global movements and a fully connected network as the fictitious force estimator.

\noindent \textbf{System Hardware} Our MID dataset is collected using the Noitom PN Studio system. Our system utilizes 6 Noitom PN Lab series sensors at a frame rate of 60fps. Our network is trained on an Nvidia RTX 4090 graphics card and is run in real-time on a laptop with Intel(R) Core(TM) i7-12700H CPU without GPU.

\begin{figure*}[t]
	\centering
 \includegraphics[width=0.9\linewidth]{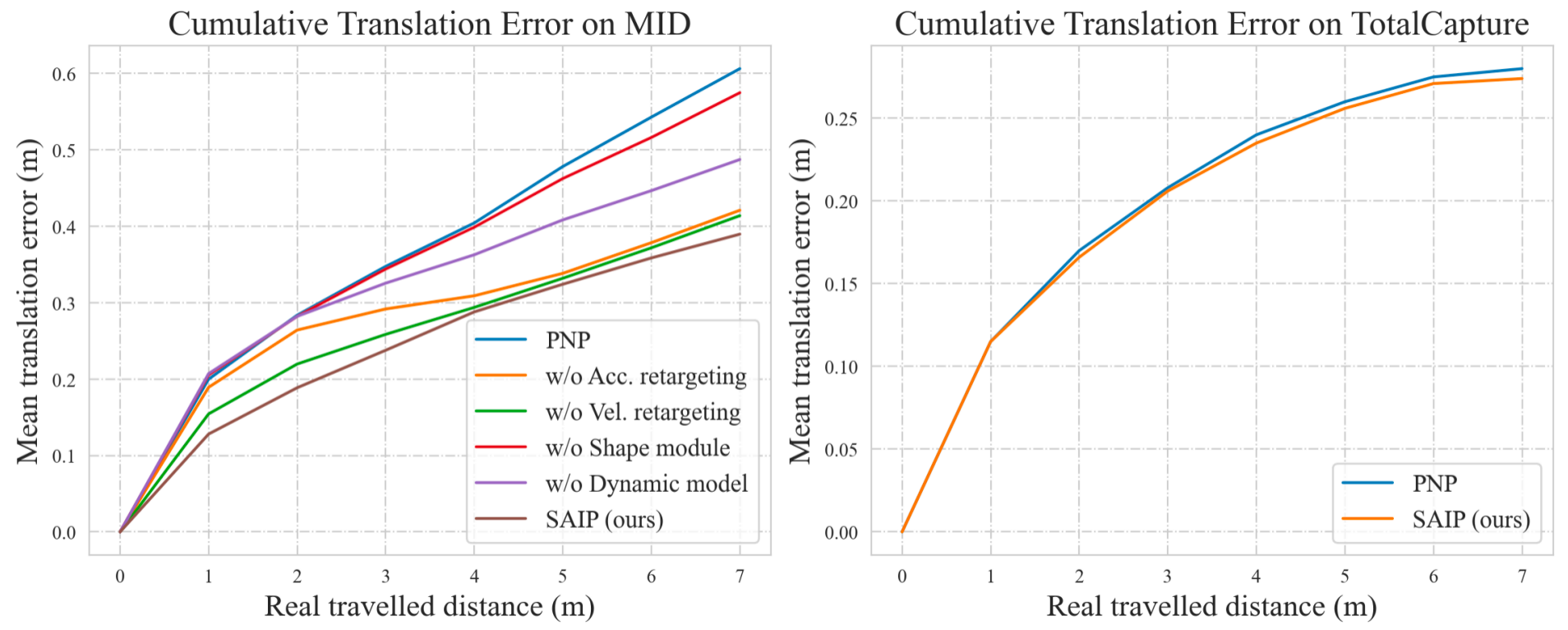}
	\caption{Comparison of translation drifting error. We plot the global position error accumulation curve with respect to the real traveled distance. A lower curve indicates smaller drift. \textbf{Left}: Ablation study on our MID dataset demonstrate the effectiveness of the proposed modules on handling subjects with diverse body shapes. \textbf{Right}: Comparison with state-of-the-art PNP on the TotalCapture dataset shows that our method achieves higher accuracy even on near-template subjects.}
	\label{fig:trans}
\end{figure*}

\begin{table}[ht]
    \centering
    \caption{Ablation studies on the kinematic signal retargeting networks, shape estimator and shape-aware physics optimization scheme. Conducted on DanceDB*.}
    \label{tab:ablation}
    \begin{tabular}{lcccccc}
        \toprule
        Method & SIP Err& Ang Err & Joint Err & Mesh Err & Jitter \\
        \midrule
        w/o $R_{acc}$ & 15.17 & 9.93 & 6.62 & 7.66 & \textbf{0.31}\\
        w/o $R_{vel}$ & 12.27 & 8.28 & 5.19 & 5.98 & 0.37\\  
        w/o Dynamic & 12.71 & 9.01 & 5.55 & 6.48 & 0.35\\
        w/o Shape & 15.82 & 10.23 & 7.20 & 8.34 & 0.36\\
        SAIP (ours) & \textbf{12.23}& \textbf{8.27}& \textbf{5.17}& \textbf{5.96}& 0.35\\
        \midrule
    \end{tabular}
\end{table}

\subsection{Comparisons}
In this section, we evaluate our proposed SAIP method against state-of-the-art inertial motion capture techniques, including TransPose ~\cite{yi2021transpose}, PIP ~\cite{yi2022physical}, TIP ~\cite{jiang2022transformer}, ASIP ~\cite{asip}, and PNP ~\cite{yi2024pnp}, using the DanceDB* dataset and real-world IMU-motion data from our MID dataset. We present the performance of our method in Tab.~\ref{tab:comparison} and our supplementary video. SAIP consistently outperforms state-of-the-art methods on both the augmented test dataset and real-world data, achieving a significant reduction in pose estimation errors compared to the second-best method. This demonstrates SAIP's robustness in motion tracking across subjects with diverse shapes, including young children. 
Notably, we retrained the transformer-based Sequence Structure Module (SSM) proposed in ASIP~\cite{asip} using the same training dataset as our model, augmented with additional child-specific data from the original adult AMASS dataset. Given transformers' ~\cite{vaswani2017attention} sensitivity to data volume, incorporating child data markedly enhanced test performance. Nevertheless, our method still achieves a 35\% improvement in mesh error. We attribute this phenomenon to the fact that simply expanding the training data is insufficient, a point we later elaborate on in the evaluation section. Although our method exhibits slightly higher jitter error than PIP, this does not indicate instability; rather, it reflects our adoption of our baseline algorithm PNP's non-inertial acceleration modeling, which increases sensitivity to motion dynamics.

\begin{figure}[t]
	\centering
 \includegraphics[width=0.9\linewidth, scale=0.8]{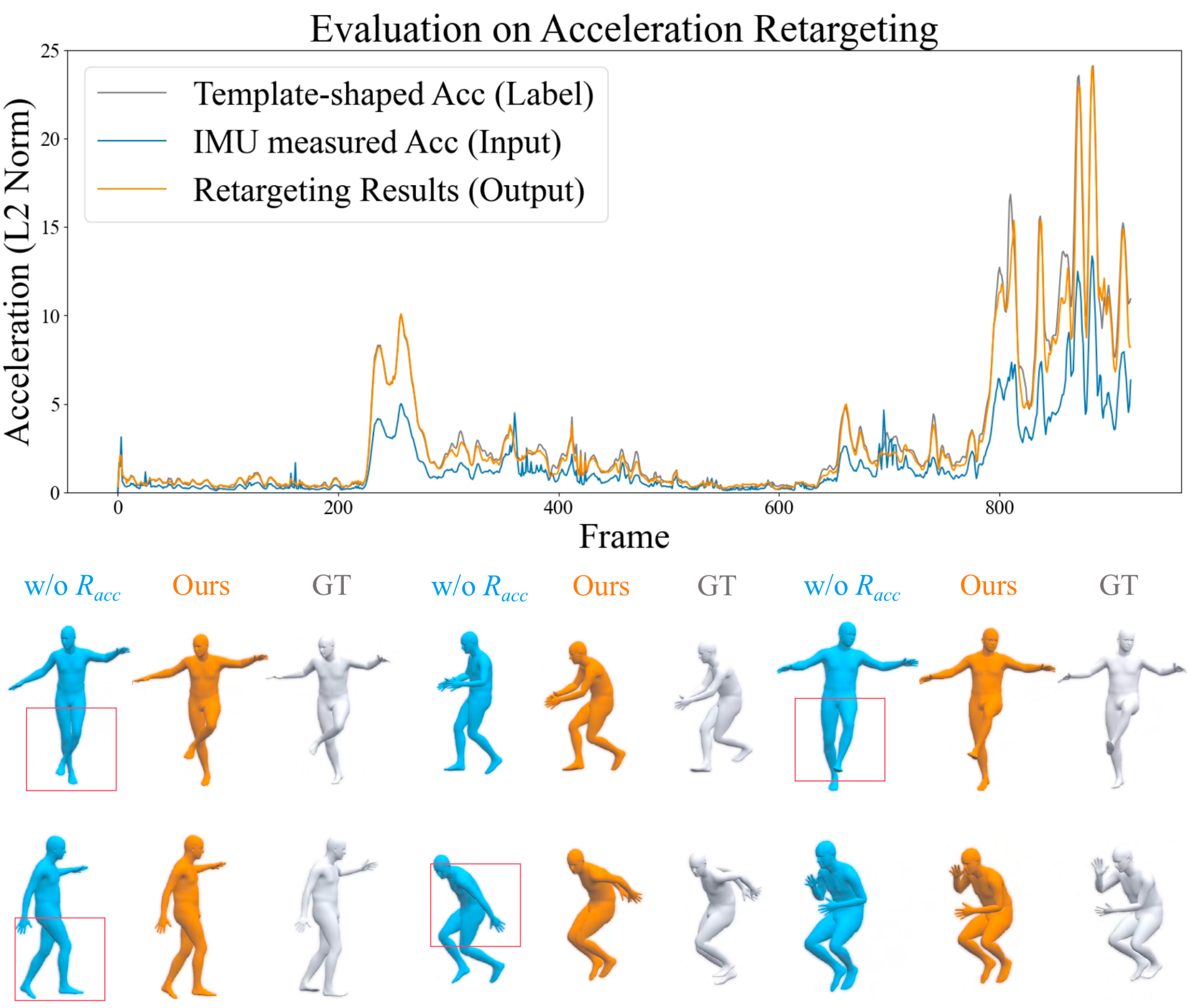}
	\caption{Demonstration of the acceleration difference caused by shape variation, and qualitative evaluation on the proposed acceleration retargeting method.}
	\label{fig:acc}
\end{figure}

\begin{figure}[t]
	\centering
 \includegraphics[width=0.9\linewidth]{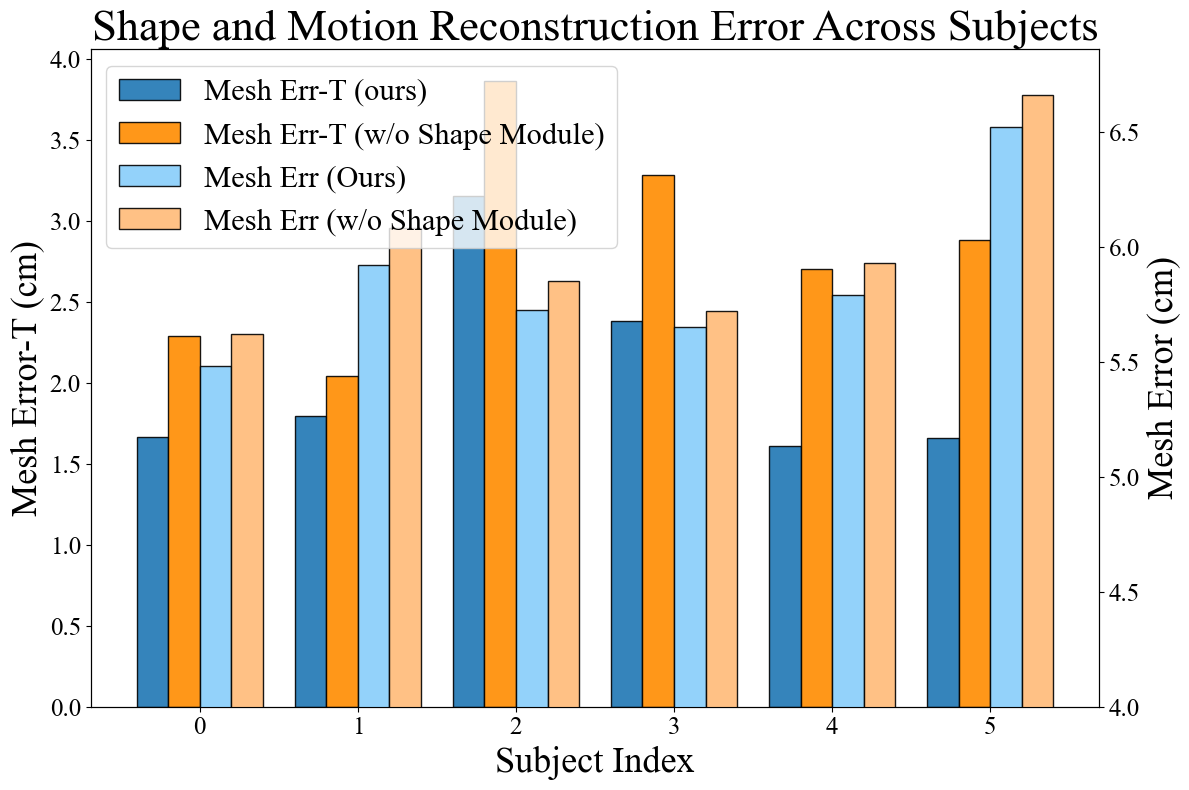}
	\caption{Shape and Motion Reconstruction Error Across Subjects. We plot the shape error in dark colors and the pose estimation error in light colors, utilizing the corresponding shapes.}
	\label{fig:bar}
\end{figure}

\subsection{Evaluations}
We conducted ablation studies on the key components of our framework, including: (1) removing $R_{acc}$ and $R_{vel}$ regression components before and after the global motion estimator, respectively, (2) using a template body shape instead of our MLP shape estimator to achieve shape awareness (w/o Shape module or w/o Shape), and (3) replacing the shape-aware physical optimization scheme with ~\cite{yi2022physical} (w/o Dynamic model or w/o Dynamic) to evaluate the effectiveness of our shape-aware optimization.

We report the quantitative results of pose accuracy in Tab. ~\ref{tab:ablation} and translation estimation error in Fig. \ref{fig:trans}, our SAIP method generally outperforms the alternatives. 
The vanilla physical optimization module in ~\cite{yi2022physical} fails to match children’s joint velocities and poses accurately, resulting in a larger translation error gap. Since $R_{vel}$ focuses on the global movement retargeting, it primarily affects global motion results and has a relatively smaller impact on pose accuracy-related metrics compared to $R_{acc}$. 
We demonstrate the gap between the accelerations of bodies with different shapes and the

\begin{table}[ht]
    \centering
    \caption{Quantitative comparison with alternative designs. }
    \label{tab:eval}
    \begin{minipage}{0.48\textwidth}
        \centering
        \setlength{\tabcolsep}{4pt}
        \begin{tabular}{c|cccc}
            \toprule
            Method & SIP Err& Ang Err & Joint Err & Mesh Err\\
            \midrule
            \multicolumn{5}{c}{w/o Retargeting nets} \\
            \midrule
            Naive Scaling & 20.25 & 12.98 & 9.10 & 10.44\\
            End-to-end Training & 17.82 & 11.92 & 9.71 & 11.34\\  
            \midrule
            \multicolumn{5}{c}{w/ Retargeting nets} \\
            \midrule
            Skeleton-only & 12.71 & 8.95 & 5.53 & 6.22\\
            Size-only & 13.01 & 9.01 & 5.55 & 6.48\\
            Shape Inference (ours) & 12.27 & 8.28 & 5.19 & 5.98\\
            Using GT Shape (ours) & \textbf{12.23}& \textbf{8.27}& \textbf{5.17}& \textbf{5.96}\\
            \bottomrule
        \end{tabular}
        \end{minipage}%
    \hfill
\end{table}

\noindent role of $R_{acc}$ in addressing this gap in Fig.~\ref{fig:acc}. 

When directly using children’s accelerations as input, the pose estimator erroneously reconstructs the pose as one resembling adults' lower-acceleration movements, such as shallow squats, limited kicking, or less pronounced limb movements. In contrast, our $R_{acc}$ effectively regresses accelerations (orange), accurately reconstructing the dynamic motions of the children.

We validated the necessity of our SAIP method by designing alternative approaches (Tab. \ref{tab:eval}). Two configurations without retargeting networks were assessed: (1) a naive scaling approach, where positional data was directly scaled based on body height as a substitute for retargeting networks (Naive Scaling), and (2) utilizing our data-augmented AMASS dataset in place of the original AMASS to train the baseline (End-to-End Training). The results underscore the critical role of independently modeling shape-conditioned IMU signals through our retargeting networks. Configurations with retargeting networks further highlighted the importance of shape awareness, where we replaced our approach with two alternative body shape representations: (3) retraining our retargeting networks using the 23 bone lengths \( B \in \mathbb{R}^{23} \) from the SMPL model (Skeleton-only), and (4) using body height alone to scale the template body (Size-only). Both alternatives, lacking detailed information on body fat and other shape proportions, performed inferior to our method. Ultimately, the results obtained using our shape prediction module (Shape Inference, ours) closely approximated those achieved with ground truth shapes (Using GT Shape, ours).

\begin{figure}[t]
	\centering
 \includegraphics[width=1\linewidth]{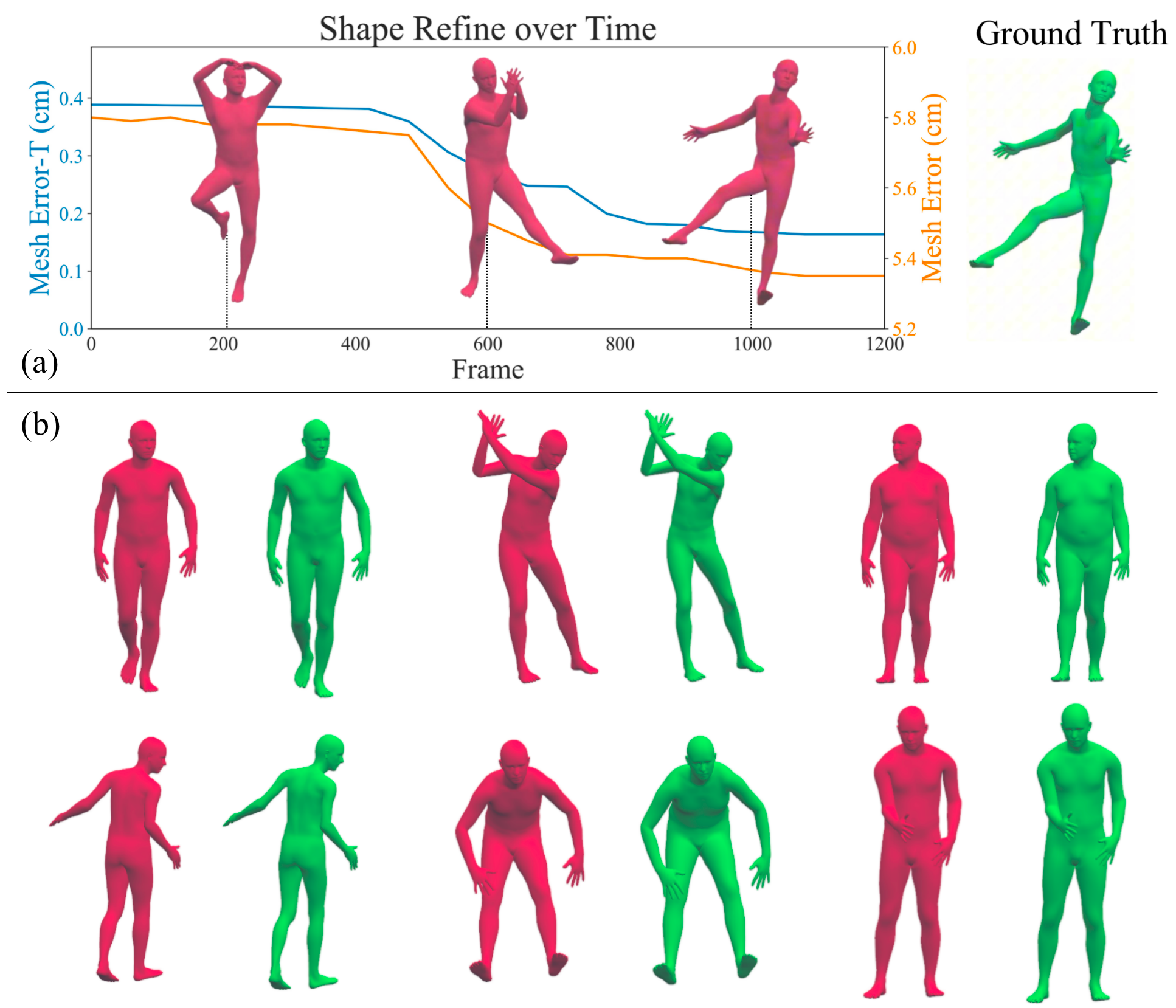}
	\caption{\textbf{(a)} Synergistically refine process: Shape-Aware Motion Tracking delivers precise kinematic data to refine the mesh, while Inertial Mesh Reconstruction enhances shape estimation, thereby further improving motion tracking accuracy. \textbf{(b)} Qualitative evaluation of Inertial Mesh Reconstruction across diverse subjects shows that the estimated shape (red) closely aligns with the ground truth (green). Examples are picked from the AMASS testset. }
	\label{fig:refine}
\end{figure}

We further evaluate the effectiveness of our MLP shape estimator. In Fig.~\ref{fig:bar}, our method (blue) pioneers shape prediction in inertial motion capture tasks, with predictions converging closer to the true shape as data accumulates. Compared to scaling a template human model (size-aware-only, orange), incorporating shape awareness also reduces the pose estimation mesh error (light-colored). In Fig.~\ref{fig:refine}, we illustrate the shape refinement process with additional qualitative results. As time progresses, the estimated shape closely aligns with the ground truth. Simultaneously, the shape estimation error (blue) decreases, leading to a corresponding reduction in motion reconstruction error (orange) using the estimated shape, thus validating our shape-pose refinement approach. Furthermore, Fig.~\ref{fig:refine} (b) demonstrates that our method successfully reconstructs diverse shapes as a byproduct, achieving strong alignment with the ground truth.

\subsection{Limitations and Discussions}

Our method addresses the limitations of baseline approaches in human motion capture by accounting for shape variations, such as those in children or individuals with diverse heights and body compositions. However, our approach has the following constraints.

Our method inherits limitations of baseline inertial motion capture techniques. For instance, it struggles to handle ground contact beyond the feet, such as crawling or rolling on the ground. Moreover, our shape-aware physical optimization scheme adopts certain assumptions from PIP \cite{yi2022physical}, such as a flat ground plane. Consequently, interactions with terrain or objects—like pull-ups or climbing—cause gravity in the optimization to pull the body back to the ground. 

In nine-axis IMUs commonly used for inertial motion capture, magnetometers calibrate yaw rotation, assuming the magnetic field aligns with the Earth's. However, in environments with magnetic interference (e.g., near powered devices or metal objects), IMU sensors experience disruptions, leading to yaw misalignment and inaccurate poses.

Our proposed method pioneers a system for predicting human body shapes using IMU sensors by modeling the relationship between shape, IMU data, and pose, enabling shape-aware pose estimation. However, this approach relies on the influence of shape variations on IMU data, making it less effective for subjects with body shapes similar to the template model, where IMU data changes are minimal and insufficient for accurate shape prediction.

The proposed SAIP system relies on human height for initializing shape information. Without this, predicting poses within acceptable error margins for individuals with significant size differences (e.g., very short children) becomes challenging, hindering the pose-shape refinement process. Additionally, our shape-aware system does not account for individuals with physical disabilities, despite the potential applications in medical rehabilitation and sports training. Future work in this area offers substantial room for improvement. Additionally, our method's shape prediction error (mesh error-T) ranges from 1.6 cm to 2.1 cm, while many optical methods based on images or videos achieve around 1.3 cm, indicating a gap in precision compared to these approaches. Future improvements in convergence accuracy are necessary to bridge this disparity.

Despite the advantages of sparse inertial motion capture, these challenges highlight significant opportunities for further research and development.

\section{Conclusion}

In this paper, we tackle the body shape factor in sparse-IMU-based human motion capture. While existing methods treat this task as modeling the correlation between IMU signals and motion, we contend that IMU signals are influenced not only by human motion but also by shape variations, leading to increased errors when applied to individuals with diverse body shapes if ignored. We propose a learning-based kinematic signal retargeting method to model shape-conditioned IMU signals, complemented by an inertial shape estimation scheme to enable shape awareness. To validate our approach, we introduce the a multi-shaped IMU-motion dataset, including pre-teen children. Experimental results validate our motivation, demonstrating that our Shape-aware Inertial Poser (SAIP) system effectively tracks motion across diverse body shapes while pioneering human shape estimation using sparse IMUs.


\begin{acks}
The authors would like to thank Xuanmiao Guo, Tianchang Chen, Luyi Fan, Xiangyi Fan, Heng Jin, Jinzhi Qian, Rongyi Quan, Ziqian Rao, JingLing Wang, Peiyuan Wang, Yile Pan and Shifan Jiang for their help on live demos and dataset collection. This work is supported by National Natural Science Foundation of China (62472364, 62072383), the Public Technology Service Platform Project of Xiamen City (No.3502Z20231043), Xiaomi Young Talents Program / Xiaomi Foundation and the Fundamental Research Funds for the Central Universities (20720240058), “Young Eagle Plan" Top Talents of Fujian Province. This work is supported by the National Key R\&D Program of China (2023YFC3305600). This work is also supported by THUIBCS, Tsinghua University, and BLBCI, Beijing Municipal Education Commission. Shihui Guo is the corresponding author. 
\end{acks}

\bibliographystyle{ACM-Reference-Format}
\bibliography{main}


\begin{thebibliography}{66}


\ifx \showCODEN    \undefined \def \showCODEN     #1{\unskip}     \fi
\ifx \showISBNx    \undefined \def \showISBNx     #1{\unskip}     \fi
\ifx \showISBNxiii \undefined \def \showISBNxiii  #1{\unskip}     \fi
\ifx \showISSN     \undefined \def \showISSN      #1{\unskip}     \fi
\ifx \showLCCN     \undefined \def \showLCCN      #1{\unskip}     \fi
\ifx \shownote     \undefined \def \shownote      #1{#1}          \fi
\ifx \showarticletitle \undefined \def \showarticletitle #1{#1}   \fi
\ifx \showURL      \undefined \def \showURL       {\relax}        \fi
\providecommand\bibfield[2]{#2}
\providecommand\bibinfo[2]{#2}
\providecommand\natexlab[1]{#1}
\providecommand\showeprint[2][]{arXiv:#2}

\bibitem[Aliakbarian et~al\mbox{.}(2022)]%
        {aliakbarian2022flag}
\bibfield{author}{\bibinfo{person}{Sadegh Aliakbarian}, \bibinfo{person}{Pashmina Cameron}, \bibinfo{person}{Federica Bogo}, \bibinfo{person}{Andrew Fitzgibbon}, {and} \bibinfo{person}{Thomas~J Cashman}.} \bibinfo{year}{2022}\natexlab{}.
\newblock \showarticletitle{Flag: Flow-based 3d avatar generation from sparse observations}. In \bibinfo{booktitle}{\emph{Proceedings of the IEEE/CVF Conference on Computer Vision and Pattern Recognition}}. \bibinfo{pages}{13253--13262}.
\newblock


\bibitem[Aloba et~al\mbox{.}(2018)]%
        {aloba2018kinder}
\bibfield{author}{\bibinfo{person}{Aishat Aloba}, \bibinfo{person}{Gianne Flores}, \bibinfo{person}{Julia Woodward}, \bibinfo{person}{Alex Shaw}, \bibinfo{person}{Amanda Castonguay}, \bibinfo{person}{Isabella Cuba}, \bibinfo{person}{Yuzhu Dong}, \bibinfo{person}{Eakta Jain}, {and} \bibinfo{person}{Lisa Anthony}.} \bibinfo{year}{2018}\natexlab{}.
\newblock \showarticletitle{Kinder-Gator: The UF Kinect Database of Child and Adult Motion.}. In \bibinfo{booktitle}{\emph{Eurographics (Short Papers)}}. \bibinfo{pages}{13--16}.
\newblock


\bibitem[Andrews et~al\mbox{.}(2016)]%
        {andrews2016real}
\bibfield{author}{\bibinfo{person}{Sheldon Andrews}, \bibinfo{person}{Ivan Huerta}, \bibinfo{person}{Taku Komura}, \bibinfo{person}{Leonid Sigal}, {and} \bibinfo{person}{Kenny Mitchell}.} \bibinfo{year}{2016}\natexlab{}.
\newblock \showarticletitle{Real-time physics-based motion capture with sparse sensors}. In \bibinfo{booktitle}{\emph{Proceedings of the 13th European conference on visual media production (CVMP 2016)}}. \bibinfo{pages}{1--10}.
\newblock


\bibitem[Aristidou et~al\mbox{.}(2019)]%
        {AMASS_DanceDB}
\bibfield{author}{\bibinfo{person}{Andreas Aristidou}, \bibinfo{person}{Ariel Shamir}, {and} \bibinfo{person}{Yiorgos Chrysanthou}.} \bibinfo{year}{2019}\natexlab{}.
\newblock \showarticletitle{Digital Dance Ethnography: {O}rganizing Large Dance Collections}.
\newblock \bibinfo{journal}{\emph{J. Comput. Cult. Herit.}} \bibinfo{volume}{12}, \bibinfo{number}{4}, Article \bibinfo{articleno}{29} (\bibinfo{date}{Nov.} \bibinfo{year}{2019}), \bibinfo{numpages}{27}~pages.
\newblock
\showISSN{1556-4673}
\href{https://doi.org/10.1145/3344383}{doi:\nolinkurl{10.1145/3344383}}


\bibitem[Bergamin et~al\mbox{.}(2019)]%
        {bergamin2019drecon}
\bibfield{author}{\bibinfo{person}{Kevin Bergamin}, \bibinfo{person}{Simon Clavet}, \bibinfo{person}{Daniel Holden}, {and} \bibinfo{person}{James~Richard Forbes}.} \bibinfo{year}{2019}\natexlab{}.
\newblock \showarticletitle{DReCon: data-driven responsive control of physics-based characters}.
\newblock \bibinfo{journal}{\emph{ACM Transactions On Graphics (TOG)}} \bibinfo{volume}{38}, \bibinfo{number}{6} (\bibinfo{year}{2019}), \bibinfo{pages}{1--11}.
\newblock


\bibitem[Black et~al\mbox{.}(2023)]%
        {black2023bedlam}
\bibfield{author}{\bibinfo{person}{Michael~J Black}, \bibinfo{person}{Priyanka Patel}, \bibinfo{person}{Joachim Tesch}, {and} \bibinfo{person}{Jinlong Yang}.} \bibinfo{year}{2023}\natexlab{}.
\newblock \showarticletitle{Bedlam: A synthetic dataset of bodies exhibiting detailed lifelike animated motion}. In \bibinfo{booktitle}{\emph{Proceedings of the IEEE/CVF Conference on Computer Vision and Pattern Recognition}}. \bibinfo{pages}{8726--8737}.
\newblock


\bibitem[Cai et~al\mbox{.}(2023)]%
        {cai2023smpler}
\bibfield{author}{\bibinfo{person}{Zhongang Cai}, \bibinfo{person}{Wanqi Yin}, \bibinfo{person}{Ailing Zeng}, \bibinfo{person}{Chen Wei}, \bibinfo{person}{Qingping Sun}, \bibinfo{person}{Wang Yanjun}, \bibinfo{person}{Hui~En Pang}, \bibinfo{person}{Haiyi Mei}, \bibinfo{person}{Mingyuan Zhang}, \bibinfo{person}{Lei Zhang}, {et~al\mbox{.}}} \bibinfo{year}{2023}\natexlab{}.
\newblock \showarticletitle{Smpler-x: Scaling up expressive human pose and shape estimation}.
\newblock \bibinfo{journal}{\emph{Advances in Neural Information Processing Systems}}  \bibinfo{volume}{36} (\bibinfo{year}{2023}), \bibinfo{pages}{11454--11468}.
\newblock


\bibitem[Chatzitofis et~al\mbox{.}(2020)]%
        {chatzitofis2020human4d}
\bibfield{author}{\bibinfo{person}{Anargyros Chatzitofis}, \bibinfo{person}{Leonidas Saroglou}, \bibinfo{person}{Prodromos Boutis}, \bibinfo{person}{Petros Drakoulis}, \bibinfo{person}{Nikolaos Zioulis}, \bibinfo{person}{Shishir Subramanyam}, \bibinfo{person}{Bart Kevelham}, \bibinfo{person}{Caecilia Charbonnier}, \bibinfo{person}{Pablo Cesar}, \bibinfo{person}{Dimitrios Zarpalas}, {et~al\mbox{.}}} \bibinfo{year}{2020}\natexlab{}.
\newblock \showarticletitle{Human4d: A human-centric multimodal dataset for motions and immersive media}.
\newblock \bibinfo{journal}{\emph{IEEE Access}}  \bibinfo{volume}{8} (\bibinfo{year}{2020}), \bibinfo{pages}{176241--176262}.
\newblock


\bibitem[Chibane et~al\mbox{.}(2020)]%
        {chibane2020implicit}
\bibfield{author}{\bibinfo{person}{Julian Chibane}, \bibinfo{person}{Thiemo Alldieck}, {and} \bibinfo{person}{Gerard Pons-Moll}.} \bibinfo{year}{2020}\natexlab{}.
\newblock \showarticletitle{Implicit functions in feature space for 3d shape reconstruction and completion}. In \bibinfo{booktitle}{\emph{Proceedings of the IEEE/CVF conference on computer vision and pattern recognition}}. \bibinfo{pages}{6970--6981}.
\newblock


\bibitem[Dittadi et~al\mbox{.}(2021)]%
        {dittadi2021full}
\bibfield{author}{\bibinfo{person}{Andrea Dittadi}, \bibinfo{person}{Sebastian Dziadzio}, \bibinfo{person}{Darren Cosker}, \bibinfo{person}{Ben Lundell}, \bibinfo{person}{Thomas~J Cashman}, {and} \bibinfo{person}{Jamie Shotton}.} \bibinfo{year}{2021}\natexlab{}.
\newblock \showarticletitle{Full-body motion from a single head-mounted device: Generating smpl poses from partial observations}. In \bibinfo{booktitle}{\emph{Proceedings of the IEEE/CVF International Conference on Computer Vision}}. \bibinfo{pages}{11687--11697}.
\newblock


\bibitem[Dong et~al\mbox{.}(2020)]%
        {dong2020adult2child}
\bibfield{author}{\bibinfo{person}{Yuzhu Dong}, \bibinfo{person}{Andreas Aristidou}, \bibinfo{person}{Ariel Shamir}, \bibinfo{person}{Moshe Mahler}, {and} \bibinfo{person}{Eakta Jain}.} \bibinfo{year}{2020}\natexlab{}.
\newblock \showarticletitle{Adult2child: Motion style transfer using cyclegans}. In \bibinfo{booktitle}{\emph{Proceedings of the 13th ACM SIGGRAPH Conference on Motion, Interaction and Games}}. \bibinfo{pages}{1--11}.
\newblock


\bibitem[Du et~al\mbox{.}(2023)]%
        {du2023agrol}
\bibfield{author}{\bibinfo{person}{Yuming Du}, \bibinfo{person}{Robin Kips}, \bibinfo{person}{Albert Pumarola}, \bibinfo{person}{Sebastian Starke}, \bibinfo{person}{Ali Thabet}, {and} \bibinfo{person}{Artsiom Sanakoyeu}.} \bibinfo{year}{2023}\natexlab{}.
\newblock \showarticletitle{Avatars Grow Legs: Generating Smooth Human Motion from Sparse Tracking Inputs with Diffusion Model}. In \bibinfo{booktitle}{\emph{CVPR}}.
\newblock


\bibitem[Featherstone(2014)]%
        {featherstone2014rigid}
\bibfield{author}{\bibinfo{person}{Roy Featherstone}.} \bibinfo{year}{2014}\natexlab{}.
\newblock \bibinfo{booktitle}{\emph{Rigid body dynamics algorithms}}.
\newblock \bibinfo{publisher}{Springer}.
\newblock


\bibitem[Felis(2016)]%
        {Felis2016}
\bibfield{author}{\bibinfo{person}{Martin~L. Felis}.} \bibinfo{year}{2016}\natexlab{}.
\newblock \showarticletitle{RBDL: an efficient rigid-body dynamics library using recursive algorithms}.
\newblock \bibinfo{journal}{\emph{Autonomous Robots}} (\bibinfo{year}{2016}), \bibinfo{pages}{1--17}.
\newblock
\showISSN{1573-7527}
\href{https://doi.org/10.1007/s10514-016-9574-0}{doi:\nolinkurl{10.1007/s10514-016-9574-0}}


\bibitem[Feng et~al\mbox{.}(2024)]%
        {feng2024stratified}
\bibfield{author}{\bibinfo{person}{Han Feng}, \bibinfo{person}{Wenchao Ma}, \bibinfo{person}{Quankai Gao}, \bibinfo{person}{Xianwei Zheng}, \bibinfo{person}{Nan Xue}, {and} \bibinfo{person}{Huijuan Xu}.} \bibinfo{year}{2024}\natexlab{}.
\newblock \showarticletitle{Stratified avatar generation from sparse observations}. In \bibinfo{booktitle}{\emph{Proceedings of the IEEE/CVF Conference on Computer Vision and Pattern Recognition}}. \bibinfo{pages}{153--163}.
\newblock


\bibitem[Huang et~al\mbox{.}(2018)]%
        {huang2018deep}
\bibfield{author}{\bibinfo{person}{Yinghao Huang}, \bibinfo{person}{Manuel Kaufmann}, \bibinfo{person}{Emre Aksan}, \bibinfo{person}{Michael~J Black}, \bibinfo{person}{Otmar Hilliges}, {and} \bibinfo{person}{Gerard Pons-Moll}.} \bibinfo{year}{2018}\natexlab{}.
\newblock \showarticletitle{Deep inertial poser: Learning to reconstruct human pose from sparse inertial measurements in real time}.
\newblock \bibinfo{journal}{\emph{ACM Transactions on Graphics (TOG)}} \bibinfo{volume}{37}, \bibinfo{number}{6} (\bibinfo{year}{2018}), \bibinfo{pages}{1--15}.
\newblock


\bibitem[Isogawa et~al\mbox{.}(2020)]%
        {isogawa2020optical}
\bibfield{author}{\bibinfo{person}{Mariko Isogawa}, \bibinfo{person}{Ye Yuan}, \bibinfo{person}{Matthew O'Toole}, {and} \bibinfo{person}{Kris~M Kitani}.} \bibinfo{year}{2020}\natexlab{}.
\newblock \showarticletitle{Optical non-line-of-sight physics-based 3d human pose estimation}. In \bibinfo{booktitle}{\emph{Proceedings of the IEEE/CVF Conference on Computer Vision and Pattern Recognition}}. \bibinfo{pages}{7013--7022}.
\newblock


\bibitem[Jiang et~al\mbox{.}(2024)]%
        {jiang2024egoposer}
\bibfield{author}{\bibinfo{person}{Jiaxi Jiang}, \bibinfo{person}{Paul Streli}, \bibinfo{person}{Manuel Meier}, {and} \bibinfo{person}{Christian Holz}.} \bibinfo{year}{2024}\natexlab{}.
\newblock \showarticletitle{Egoposer: Robust real-time egocentric pose estimation from sparse and intermittent observations everywhere}. In \bibinfo{booktitle}{\emph{European Conference on Computer Vision}}. Springer, \bibinfo{pages}{277--294}.
\newblock


\bibitem[Jiang et~al\mbox{.}(2022a)]%
        {Jiang2022AvatarPoserAF}
\bibfield{author}{\bibinfo{person}{Jiaxi Jiang}, \bibinfo{person}{Paul Streli}, \bibinfo{person}{Huajian Qiu}, \bibinfo{person}{Andreas~Rene Fender}, \bibinfo{person}{Larissa Laich}, \bibinfo{person}{Patrick Snape}, {and} \bibinfo{person}{Christian Holz}.} \bibinfo{year}{2022}\natexlab{a}.
\newblock \showarticletitle{AvatarPoser: Articulated Full-Body Pose Tracking from Sparse Motion Sensing}. In \bibinfo{booktitle}{\emph{European Conference on Computer Vision}}.
\newblock
\urldef\tempurl%
\url{https://api.semanticscholar.org/CorpusID:251135349}
\showURL{%
\tempurl}


\bibitem[Jiang et~al\mbox{.}(2022b)]%
        {jiang2022transformer}
\bibfield{author}{\bibinfo{person}{Yifeng Jiang}, \bibinfo{person}{Yuting Ye}, \bibinfo{person}{Deepak Gopinath}, \bibinfo{person}{Jungdam Won}, \bibinfo{person}{Alexander~W Winkler}, {and} \bibinfo{person}{C~Karen Liu}.} \bibinfo{year}{2022}\natexlab{b}.
\newblock \showarticletitle{Transformer inertial poser: Real-time human motion reconstruction from sparse imus with simultaneous terrain generation}. In \bibinfo{booktitle}{\emph{SIGGRAPH Asia 2022 Conference Papers}}. \bibinfo{pages}{1--9}.
\newblock


\bibitem[Kocabas et~al\mbox{.}(2020)]%
        {kocabas2020vibe}
\bibfield{author}{\bibinfo{person}{Muhammed Kocabas}, \bibinfo{person}{Nikos Athanasiou}, {and} \bibinfo{person}{Michael~J Black}.} \bibinfo{year}{2020}\natexlab{}.
\newblock \showarticletitle{Vibe: Video inference for human body pose and shape estimation}. In \bibinfo{booktitle}{\emph{Proceedings of the IEEE/CVF conference on computer vision and pattern recognition}}. \bibinfo{pages}{5253--5263}.
\newblock


\bibitem[Kratzer et~al\mbox{.}(2020)]%
        {kratzer2020mogaze}
\bibfield{author}{\bibinfo{person}{Philipp Kratzer}, \bibinfo{person}{Simon Bihlmaier}, \bibinfo{person}{Niteesh~Balachandra Midlagajni}, \bibinfo{person}{Rohit Prakash}, \bibinfo{person}{Marc Toussaint}, {and} \bibinfo{person}{Jim Mainprice}.} \bibinfo{year}{2020}\natexlab{}.
\newblock \showarticletitle{Mogaze: A dataset of full-body motions that includes workspace geometry and eye-gaze}.
\newblock \bibinfo{journal}{\emph{IEEE Robotics and Automation Letters}} \bibinfo{volume}{6}, \bibinfo{number}{2} (\bibinfo{year}{2020}), \bibinfo{pages}{367--373}.
\newblock


\bibitem[Lee and Joo(2024)]%
        {lee2024mocap}
\bibfield{author}{\bibinfo{person}{Jiye Lee} {and} \bibinfo{person}{Hanbyul Joo}.} \bibinfo{year}{2024}\natexlab{}.
\newblock \showarticletitle{Mocap everyone everywhere: Lightweight motion capture with smartwatches and a head-mounted camera}. In \bibinfo{booktitle}{\emph{Proceedings of the IEEE/CVF conference on computer vision and pattern recognition}}. \bibinfo{pages}{1091--1100}.
\newblock


\bibitem[Lee et~al\mbox{.}(2023)]%
        {lee2023questenvsim}
\bibfield{author}{\bibinfo{person}{Sunmin Lee}, \bibinfo{person}{Sebastian Starke}, \bibinfo{person}{Yuting Ye}, \bibinfo{person}{Jungdam Won}, {and} \bibinfo{person}{Alexander Winkler}.} \bibinfo{year}{2023}\natexlab{}.
\newblock \showarticletitle{Questenvsim: Environment-aware simulated motion tracking from sparse sensors}. In \bibinfo{booktitle}{\emph{ACM SIGGRAPH 2023 Conference Proceedings}}. \bibinfo{pages}{1--9}.
\newblock


\bibitem[Li et~al\mbox{.}(2021b)]%
        {li2021hybrik}
\bibfield{author}{\bibinfo{person}{Jiefeng Li}, \bibinfo{person}{Chao Xu}, \bibinfo{person}{Zhicun Chen}, \bibinfo{person}{Siyuan Bian}, \bibinfo{person}{Lixin Yang}, {and} \bibinfo{person}{Cewu Lu}.} \bibinfo{year}{2021}\natexlab{b}.
\newblock \showarticletitle{Hybrik: A hybrid analytical-neural inverse kinematics solution for 3d human pose and shape estimation}. In \bibinfo{booktitle}{\emph{Proceedings of the IEEE/CVF conference on computer vision and pattern recognition}}. \bibinfo{pages}{3383--3393}.
\newblock


\bibitem[Li et~al\mbox{.}(2021a)]%
        {li20213d}
\bibfield{author}{\bibinfo{person}{Zhongguo Li}, \bibinfo{person}{Magnus Oskarsson}, {and} \bibinfo{person}{Anders Heyden}.} \bibinfo{year}{2021}\natexlab{a}.
\newblock \showarticletitle{3d human pose and shape estimation through collaborative learning and multi-view model-fitting}. In \bibinfo{booktitle}{\emph{Proceedings of the IEEE/CVF winter conference on applications of computer vision}}. \bibinfo{pages}{1888--1897}.
\newblock


\bibitem[Li et~al\mbox{.}(2019)]%
        {li2019estimating}
\bibfield{author}{\bibinfo{person}{Zongmian Li}, \bibinfo{person}{Jiri Sedlar}, \bibinfo{person}{Justin Carpentier}, \bibinfo{person}{Ivan Laptev}, \bibinfo{person}{Nicolas Mansard}, {and} \bibinfo{person}{Josef Sivic}.} \bibinfo{year}{2019}\natexlab{}.
\newblock \showarticletitle{Estimating 3d motion and forces of person-object interactions from monocular video}. In \bibinfo{booktitle}{\emph{Proceedings of the IEEE/CVF Conference on Computer Vision and Pattern Recognition}}. \bibinfo{pages}{8640--8649}.
\newblock


\bibitem[Liang et~al\mbox{.}(2023)]%
        {liang2023hybridcap}
\bibfield{author}{\bibinfo{person}{Han Liang}, \bibinfo{person}{Yannan He}, \bibinfo{person}{Chengfeng Zhao}, \bibinfo{person}{Mutian Li}, \bibinfo{person}{Jingya Wang}, \bibinfo{person}{Jingyi Yu}, {and} \bibinfo{person}{Lan Xu}.} \bibinfo{year}{2023}\natexlab{}.
\newblock \showarticletitle{Hybridcap: Inertia-aid monocular capture of challenging human motions}. In \bibinfo{booktitle}{\emph{Proceedings of the AAAI conference on artificial intelligence}}, Vol.~\bibinfo{volume}{37}. \bibinfo{pages}{1539--1548}.
\newblock


\bibitem[Lin et~al\mbox{.}(2023)]%
        {lin2023motion}
\bibfield{author}{\bibinfo{person}{Jing Lin}, \bibinfo{person}{Ailing Zeng}, \bibinfo{person}{Shunlin Lu}, \bibinfo{person}{Yuanhao Cai}, \bibinfo{person}{Ruimao Zhang}, \bibinfo{person}{Haoqian Wang}, {and} \bibinfo{person}{Lei Zhang}.} \bibinfo{year}{2023}\natexlab{}.
\newblock \showarticletitle{Motion-x: A large-scale 3d expressive whole-body human motion dataset}.
\newblock \bibinfo{journal}{\emph{Advances in Neural Information Processing Systems}}  \bibinfo{volume}{36} (\bibinfo{year}{2023}), \bibinfo{pages}{25268--25280}.
\newblock


\bibitem[Loper et~al\mbox{.}(2023)]%
        {loper2023smpl}
\bibfield{author}{\bibinfo{person}{Matthew Loper}, \bibinfo{person}{Naureen Mahmood}, \bibinfo{person}{Javier Romero}, \bibinfo{person}{Gerard Pons-Moll}, {and} \bibinfo{person}{Michael~J Black}.} \bibinfo{year}{2023}\natexlab{}.
\newblock \showarticletitle{SMPL: A skinned multi-person linear model}.
\newblock In \bibinfo{booktitle}{\emph{Seminal Graphics Papers: Pushing the Boundaries, Volume 2}}. \bibinfo{pages}{851--866}.
\newblock


\bibitem[Ma et~al\mbox{.}(2024)]%
        {ma2024nymeria}
\bibfield{author}{\bibinfo{person}{Lingni Ma}, \bibinfo{person}{Yuting Ye}, \bibinfo{person}{Fangzhou Hong}, \bibinfo{person}{Vladimir Guzov}, \bibinfo{person}{Yifeng Jiang}, \bibinfo{person}{Rowan Postyeni}, \bibinfo{person}{Luis Pesqueira}, \bibinfo{person}{Alexander Gamino}, \bibinfo{person}{Vijay Baiyya}, \bibinfo{person}{Hyo~Jin Kim}, {et~al\mbox{.}}} \bibinfo{year}{2024}\natexlab{}.
\newblock \showarticletitle{Nymeria: A massive collection of multimodal egocentric daily motion in the wild}. In \bibinfo{booktitle}{\emph{European Conference on Computer Vision}}. Springer, \bibinfo{pages}{445--465}.
\newblock


\bibitem[Mahmood et~al\mbox{.}(2019)]%
        {AMASS}
\bibfield{author}{\bibinfo{person}{Naureen Mahmood}, \bibinfo{person}{Nima Ghorbani}, \bibinfo{person}{Nikolaus~F. Troje}, \bibinfo{person}{Gerard Pons-Moll}, {and} \bibinfo{person}{Michael~J. Black}.} \bibinfo{year}{2019}\natexlab{}.
\newblock \showarticletitle{{AMASS}: Archive of Motion Capture as Surface Shapes}. In \bibinfo{booktitle}{\emph{International Conference on Computer Vision}}. \bibinfo{pages}{5442--5451}.
\newblock


\bibitem[Maurice et~al\mbox{.}(2019)]%
        {maurice2019human}
\bibfield{author}{\bibinfo{person}{Pauline Maurice}, \bibinfo{person}{Adrien Malais{\'e}}, \bibinfo{person}{Cl{\'e}lie Amiot}, \bibinfo{person}{Nicolas Paris}, \bibinfo{person}{Guy-Junior Richard}, \bibinfo{person}{Olivier Rochel}, {and} \bibinfo{person}{Serena Ivaldi}.} \bibinfo{year}{2019}\natexlab{}.
\newblock \showarticletitle{Human movement and ergonomics: An industry-oriented dataset for collaborative robotics}.
\newblock \bibinfo{journal}{\emph{The International Journal of Robotics Research}} \bibinfo{volume}{38}, \bibinfo{number}{14} (\bibinfo{year}{2019}), \bibinfo{pages}{1529--1537}.
\newblock


\bibitem[Noitom(2017)]%
        {noitom2017perception}
\bibfield{author}{\bibinfo{person}{Noitom}.} \bibinfo{year}{2017}\natexlab{}.
\newblock \showarticletitle{Perception neuron}.
\newblock  (\bibinfo{year}{2017}).
\newblock
\urldef\tempurl%
\url{https://www.noitom.com/}
\showURL{%
\tempurl}


\bibitem[Palermo et~al\mbox{.}(2022)]%
        {palermo2022complete}
\bibfield{author}{\bibinfo{person}{Manuel Palermo}, \bibinfo{person}{Sara Cerqueira}, \bibinfo{person}{Jo{\~a}o Andr{\'e}}, \bibinfo{person}{Ant{\'o}nio Pereira}, {and} \bibinfo{person}{Cristina~P Santos}.} \bibinfo{year}{2022}\natexlab{}.
\newblock \showarticletitle{Complete Inertial Pose Dataset: from raw measurements to pose with low-cost and high-end MARG sensors}.
\newblock \bibinfo{journal}{\emph{arXiv preprint arXiv:2202.06164}} (\bibinfo{year}{2022}).
\newblock


\bibitem[Pang et~al\mbox{.}(2022)]%
        {pang2022benchmarking}
\bibfield{author}{\bibinfo{person}{Hui~En Pang}, \bibinfo{person}{Zhongang Cai}, \bibinfo{person}{Lei Yang}, \bibinfo{person}{Tianwei Zhang}, {and} \bibinfo{person}{Ziwei Liu}.} \bibinfo{year}{2022}\natexlab{}.
\newblock \showarticletitle{Benchmarking and analyzing 3d human pose and shape estimation beyond algorithms}.
\newblock \bibinfo{journal}{\emph{Advances in Neural Information Processing Systems}}  \bibinfo{volume}{35} (\bibinfo{year}{2022}), \bibinfo{pages}{26034--26051}.
\newblock


\bibitem[Paulich et~al\mbox{.}(2018)]%
        {paulich2018xsens}
\bibfield{author}{\bibinfo{person}{Monique Paulich}, \bibinfo{person}{Martin Schepers}, \bibinfo{person}{Nina Rudigkeit}, {and} \bibinfo{person}{Giovanni Bellusci}.} \bibinfo{year}{2018}\natexlab{}.
\newblock \showarticletitle{Xsens MTw Awinda: Miniature wireless inertial-magnetic motion tracker for highly accurate 3D kinematic applications}.
\newblock \bibinfo{journal}{\emph{Xsens: Enschede, The Netherlands}} (\bibinfo{year}{2018}), \bibinfo{pages}{1--9}.
\newblock


\bibitem[Plappert et~al\mbox{.}(2016)]%
        {plappert2016kit}
\bibfield{author}{\bibinfo{person}{Matthias Plappert}, \bibinfo{person}{Christian Mandery}, {and} \bibinfo{person}{Tamim Asfour}.} \bibinfo{year}{2016}\natexlab{}.
\newblock \showarticletitle{The kit motion-language dataset}.
\newblock \bibinfo{journal}{\emph{Big data}} \bibinfo{volume}{4}, \bibinfo{number}{4} (\bibinfo{year}{2016}), \bibinfo{pages}{236--252}.
\newblock


\bibitem[Point(2011)]%
        {point2011optitrack}
\bibfield{author}{\bibinfo{person}{Natural Point}.} \bibinfo{year}{2011}\natexlab{}.
\newblock \showarticletitle{Optitrack. Natural Point, Inc.}
\newblock \bibinfo{journal}{\emph{Natural Point Inc}} (\bibinfo{year}{2011}).
\newblock


\bibitem[Ponton et~al\mbox{.}(2023)]%
        {ponton2023sparseposer}
\bibfield{author}{\bibinfo{person}{Jose~Luis Ponton}, \bibinfo{person}{Haoran Yun}, \bibinfo{person}{Andreas Aristidou}, \bibinfo{person}{Carlos Andujar}, {and} \bibinfo{person}{Nuria Pelechano}.} \bibinfo{year}{2023}\natexlab{}.
\newblock \showarticletitle{SparsePoser: Real-time full-body motion reconstruction from sparse data}.
\newblock \bibinfo{journal}{\emph{ACM Transactions on Graphics}} \bibinfo{volume}{43}, \bibinfo{number}{1} (\bibinfo{year}{2023}), \bibinfo{pages}{1--14}.
\newblock


\bibitem[Rempe et~al\mbox{.}(2020)]%
        {rempe2020contact}
\bibfield{author}{\bibinfo{person}{Davis Rempe}, \bibinfo{person}{Leonidas~J Guibas}, \bibinfo{person}{Aaron Hertzmann}, \bibinfo{person}{Bryan Russell}, \bibinfo{person}{Ruben Villegas}, {and} \bibinfo{person}{Jimei Yang}.} \bibinfo{year}{2020}\natexlab{}.
\newblock \showarticletitle{Contact and human dynamics from monocular video}. In \bibinfo{booktitle}{\emph{Computer Vision--ECCV 2020: 16th European Conference, Glasgow, UK, August 23--28, 2020, Proceedings, Part V 16}}. Springer, \bibinfo{pages}{71--87}.
\newblock


\bibitem[Schreiner et~al\mbox{.}(2024)]%
        {schreiner2024adapt}
\bibfield{author}{\bibinfo{person}{Paul Schreiner}, \bibinfo{person}{Rasmus Netterstr{\o}m}, \bibinfo{person}{Hang Yin}, \bibinfo{person}{Sune Darkner}, {and} \bibinfo{person}{Kenny Erleben}.} \bibinfo{year}{2024}\natexlab{}.
\newblock \showarticletitle{ADAPT: AI-Driven Artefact Purging Technique for IMU Based Motion Capture}. In \bibinfo{booktitle}{\emph{Computer Graphics Forum}}, Vol.~\bibinfo{volume}{43}. Wiley Online Library, \bibinfo{pages}{e15172}.
\newblock


\bibitem[Sengupta et~al\mbox{.}(2020)]%
        {sengupta2020synthetic}
\bibfield{author}{\bibinfo{person}{Akash Sengupta}, \bibinfo{person}{Ignas Budvytis}, {and} \bibinfo{person}{Roberto Cipolla}.} \bibinfo{year}{2020}\natexlab{}.
\newblock \showarticletitle{Synthetic training for accurate 3d human pose and shape estimation in the wild}.
\newblock \bibinfo{journal}{\emph{arXiv preprint arXiv:2009.10013}} (\bibinfo{year}{2020}).
\newblock


\bibitem[Shen et~al\mbox{.}(2023)]%
        {shen2023global}
\bibfield{author}{\bibinfo{person}{Xiaolong Shen}, \bibinfo{person}{Zongxin Yang}, \bibinfo{person}{Xiaohan Wang}, \bibinfo{person}{Jianxin Ma}, \bibinfo{person}{Chang Zhou}, {and} \bibinfo{person}{Yi Yang}.} \bibinfo{year}{2023}\natexlab{}.
\newblock \showarticletitle{Global-to-local modeling for video-based 3d human pose and shape estimation}. In \bibinfo{booktitle}{\emph{Proceedings of the IEEE/CVF Conference on Computer Vision and Pattern Recognition}}. \bibinfo{pages}{8887--8896}.
\newblock


\bibitem[Shimada et~al\mbox{.}(2020)]%
        {shimada2020physcap}
\bibfield{author}{\bibinfo{person}{Soshi Shimada}, \bibinfo{person}{Vladislav Golyanik}, \bibinfo{person}{Weipeng Xu}, {and} \bibinfo{person}{Christian Theobalt}.} \bibinfo{year}{2020}\natexlab{}.
\newblock \showarticletitle{Physcap: Physically plausible monocular 3d motion capture in real time}.
\newblock \bibinfo{journal}{\emph{ACM Transactions on Graphics (ToG)}} \bibinfo{volume}{39}, \bibinfo{number}{6} (\bibinfo{year}{2020}), \bibinfo{pages}{1--16}.
\newblock


\bibitem[Sun et~al\mbox{.}(2019)]%
        {sun2019deep}
\bibfield{author}{\bibinfo{person}{Ke Sun}, \bibinfo{person}{Bin Xiao}, \bibinfo{person}{Dong Liu}, {and} \bibinfo{person}{Jingdong Wang}.} \bibinfo{year}{2019}\natexlab{}.
\newblock \showarticletitle{Deep high-resolution representation learning for human pose estimation}. In \bibinfo{booktitle}{\emph{Proceedings of the IEEE/CVF conference on computer vision and pattern recognition}}. \bibinfo{pages}{5693--5703}.
\newblock


\bibitem[Trumble et~al\mbox{.}(2017)]%
        {trumble2017total}
\bibfield{author}{\bibinfo{person}{Matthew Trumble}, \bibinfo{person}{Andrew Gilbert}, \bibinfo{person}{Charles Malleson}, \bibinfo{person}{Adrian Hilton}, {and} \bibinfo{person}{John~P Collomosse}.} \bibinfo{year}{2017}\natexlab{}.
\newblock \showarticletitle{Total capture: 3D human pose estimation fusing video and inertial sensors.}. In \bibinfo{booktitle}{\emph{BMVC}}, Vol.~\bibinfo{volume}{2}. London, UK, \bibinfo{pages}{1--13}.
\newblock


\bibitem[Vaswani et~al\mbox{.}(2017)]%
        {vaswani2017attention}
\bibfield{author}{\bibinfo{person}{Ashish Vaswani}, \bibinfo{person}{Noam Shazeer}, \bibinfo{person}{Niki Parmar}, \bibinfo{person}{Jakob Uszkoreit}, \bibinfo{person}{Llion Jones}, \bibinfo{person}{Aidan~N Gomez}, \bibinfo{person}{{\L}ukasz Kaiser}, {and} \bibinfo{person}{Illia Polosukhin}.} \bibinfo{year}{2017}\natexlab{}.
\newblock \showarticletitle{Attention is all you need}.
\newblock \bibinfo{journal}{\emph{Advances in neural information processing systems}}  \bibinfo{volume}{30} (\bibinfo{year}{2017}).
\newblock


\bibitem[Von~Marcard et~al\mbox{.}(2017)]%
        {von2017sparse}
\bibfield{author}{\bibinfo{person}{Timo Von~Marcard}, \bibinfo{person}{Bodo Rosenhahn}, \bibinfo{person}{Michael~J Black}, {and} \bibinfo{person}{Gerard Pons-Moll}.} \bibinfo{year}{2017}\natexlab{}.
\newblock \showarticletitle{Sparse inertial poser: Automatic 3d human pose estimation from sparse imus}. In \bibinfo{booktitle}{\emph{Computer graphics forum}}, Vol.~\bibinfo{volume}{36}. Wiley Online Library, \bibinfo{pages}{349--360}.
\newblock


\bibitem[Winkler et~al\mbox{.}(2022)]%
        {winkler2022questsim}
\bibfield{author}{\bibinfo{person}{Alexander Winkler}, \bibinfo{person}{Jungdam Won}, {and} \bibinfo{person}{Yuting Ye}.} \bibinfo{year}{2022}\natexlab{}.
\newblock \showarticletitle{Questsim: Human motion tracking from sparse sensors with simulated avatars}. In \bibinfo{booktitle}{\emph{SIGGRAPH Asia 2022 Conference Papers}}. \bibinfo{pages}{1--8}.
\newblock


\bibitem[Wu et~al\mbox{.}(2024)]%
        {asip}
\bibfield{author}{\bibinfo{person}{Yinghao Wu}, \bibinfo{person}{Chaoran Wang}, \bibinfo{person}{Lu Yin}, \bibinfo{person}{Shihui Guo}, {and} \bibinfo{person}{Yipeng Qin}.} \bibinfo{year}{2024}\natexlab{}.
\newblock \showarticletitle{Accurate and Steady Inertial Pose Estimation through Sequence Structure Learning and Modulation}. In \bibinfo{booktitle}{\emph{NeurIPS}}.
\newblock


\bibitem[Yang et~al\mbox{.}(2024)]%
        {yang2024divatrack}
\bibfield{author}{\bibinfo{person}{Dongseok Yang}, \bibinfo{person}{Jiho Kang}, \bibinfo{person}{Lingni Ma}, \bibinfo{person}{Joseph Greer}, \bibinfo{person}{Yuting Ye}, {and} \bibinfo{person}{Sung-Hee Lee}.} \bibinfo{year}{2024}\natexlab{}.
\newblock \showarticletitle{DivaTrack: Diverse Bodies and Motions from Acceleration-Enhanced Three-Point Trackers}. In \bibinfo{booktitle}{\emph{Computer Graphics Forum}}, Vol.~\bibinfo{volume}{43}. Wiley Online Library, \bibinfo{pages}{e15057}.
\newblock


\bibitem[Yang et~al\mbox{.}(2021a)]%
        {yang2021lobstr}
\bibfield{author}{\bibinfo{person}{Dongseok Yang}, \bibinfo{person}{Doyeon Kim}, {and} \bibinfo{person}{Sung-Hee Lee}.} \bibinfo{year}{2021}\natexlab{a}.
\newblock \showarticletitle{Lobstr: Real-time lower-body pose prediction from sparse upper-body tracking signals}. In \bibinfo{booktitle}{\emph{Computer Graphics Forum}}, Vol.~\bibinfo{volume}{40}. Wiley Online Library, \bibinfo{pages}{265--275}.
\newblock


\bibitem[Yang et~al\mbox{.}(2021b)]%
        {yang2021viser}
\bibfield{author}{\bibinfo{person}{Gengshan Yang}, \bibinfo{person}{Deqing Sun}, \bibinfo{person}{Varun Jampani}, \bibinfo{person}{Daniel Vlasic}, \bibinfo{person}{Forrester Cole}, \bibinfo{person}{Ce Liu}, {and} \bibinfo{person}{Deva Ramanan}.} \bibinfo{year}{2021}\natexlab{b}.
\newblock \showarticletitle{Viser: Video-specific surface embeddings for articulated 3d shape reconstruction}.
\newblock \bibinfo{journal}{\emph{Advances in Neural Information Processing Systems}}  \bibinfo{volume}{34} (\bibinfo{year}{2021}), \bibinfo{pages}{19326--19338}.
\newblock


\bibitem[Ye et~al\mbox{.}(2022)]%
        {ye2022faster}
\bibfield{author}{\bibinfo{person}{Hang Ye}, \bibinfo{person}{Wentao Zhu}, \bibinfo{person}{Chunyu Wang}, \bibinfo{person}{Rujie Wu}, {and} \bibinfo{person}{Yizhou Wang}.} \bibinfo{year}{2022}\natexlab{}.
\newblock \showarticletitle{Faster voxelpose: Real-time 3d human pose estimation by orthographic projection}. In \bibinfo{booktitle}{\emph{European Conference on Computer Vision}}. Springer, \bibinfo{pages}{142--159}.
\newblock


\bibitem[Yi et~al\mbox{.}(2022)]%
        {yi2022physical}
\bibfield{author}{\bibinfo{person}{Xinyu Yi}, \bibinfo{person}{Yuxiao Zhou}, \bibinfo{person}{Marc Habermann}, \bibinfo{person}{Soshi Shimada}, \bibinfo{person}{Vladislav Golyanik}, \bibinfo{person}{Christian Theobalt}, {and} \bibinfo{person}{Feng Xu}.} \bibinfo{year}{2022}\natexlab{}.
\newblock \showarticletitle{Physical inertial poser (pip): Physics-aware real-time human motion tracking from sparse inertial sensors}. In \bibinfo{booktitle}{\emph{Proceedings of the IEEE/CVF Conference on Computer Vision and Pattern Recognition}}. \bibinfo{pages}{13167--13178}.
\newblock


\bibitem[Yi et~al\mbox{.}(2021)]%
        {yi2021transpose}
\bibfield{author}{\bibinfo{person}{Xinyu Yi}, \bibinfo{person}{Yuxiao Zhou}, {and} \bibinfo{person}{Feng Xu}.} \bibinfo{year}{2021}\natexlab{}.
\newblock \showarticletitle{Transpose: Real-time 3d human translation and pose estimation with six inertial sensors}.
\newblock \bibinfo{journal}{\emph{ACM Transactions on Graphics (TOG)}} \bibinfo{volume}{40}, \bibinfo{number}{4} (\bibinfo{year}{2021}), \bibinfo{pages}{1--13}.
\newblock


\bibitem[Yi et~al\mbox{.}(2024)]%
        {yi2024pnp}
\bibfield{author}{\bibinfo{person}{Xinyu Yi}, \bibinfo{person}{Yuxiao Zhou}, {and} \bibinfo{person}{Feng Xu}.} \bibinfo{year}{2024}\natexlab{}.
\newblock \showarticletitle{Physical Non-inertial Poser (PNP): Modeling Non-inertial Effects in Sparse-inertial Human Motion Capture}. In \bibinfo{booktitle}{\emph{SIGGRAPH 2024 Conference Papers}}.
\newblock


\bibitem[Yu et~al\mbox{.}(2021a)]%
        {yu2021human}
\bibfield{author}{\bibinfo{person}{Ri Yu}, \bibinfo{person}{Hwangpil Park}, {and} \bibinfo{person}{Jehee Lee}.} \bibinfo{year}{2021}\natexlab{a}.
\newblock \showarticletitle{Human dynamics from monocular video with dynamic camera movements}.
\newblock \bibinfo{journal}{\emph{ACM Transactions on Graphics (TOG)}} \bibinfo{volume}{40}, \bibinfo{number}{6} (\bibinfo{year}{2021}), \bibinfo{pages}{1--14}.
\newblock


\bibitem[Yu et~al\mbox{.}(2021b)]%
        {yu2021function4d}
\bibfield{author}{\bibinfo{person}{Tao Yu}, \bibinfo{person}{Zerong Zheng}, \bibinfo{person}{Kaiwen Guo}, \bibinfo{person}{Pengpeng Liu}, \bibinfo{person}{Qionghai Dai}, {and} \bibinfo{person}{Yebin Liu}.} \bibinfo{year}{2021}\natexlab{b}.
\newblock \showarticletitle{Function4d: Real-time human volumetric capture from very sparse consumer rgbd sensors}. In \bibinfo{booktitle}{\emph{Proceedings of the IEEE/CVF conference on computer vision and pattern recognition}}. \bibinfo{pages}{5746--5756}.
\newblock


\bibitem[Yuan and Kitani(2019)]%
        {yuan2019ego}
\bibfield{author}{\bibinfo{person}{Ye Yuan} {and} \bibinfo{person}{Kris Kitani}.} \bibinfo{year}{2019}\natexlab{}.
\newblock \showarticletitle{Ego-pose estimation and forecasting as real-time pd control}. In \bibinfo{booktitle}{\emph{Proceedings of the IEEE/CVF International Conference on Computer Vision}}. \bibinfo{pages}{10082--10092}.
\newblock


\bibitem[Yuan et~al\mbox{.}(2021)]%
        {yuan2021simpoe}
\bibfield{author}{\bibinfo{person}{Ye Yuan}, \bibinfo{person}{Shih-En Wei}, \bibinfo{person}{Tomas Simon}, \bibinfo{person}{Kris Kitani}, {and} \bibinfo{person}{Jason Saragih}.} \bibinfo{year}{2021}\natexlab{}.
\newblock \showarticletitle{Simpoe: Simulated character control for 3d human pose estimation}. In \bibinfo{booktitle}{\emph{Proceedings of the IEEE/CVF conference on computer vision and pattern recognition}}. \bibinfo{pages}{7159--7169}.
\newblock


\bibitem[Zhang et~al\mbox{.}(2021)]%
        {zhang2021direct}
\bibfield{author}{\bibinfo{person}{Jianfeng Zhang}, \bibinfo{person}{Yujun Cai}, \bibinfo{person}{Shuicheng Yan}, \bibinfo{person}{Jiashi Feng}, {et~al\mbox{.}}} \bibinfo{year}{2021}\natexlab{}.
\newblock \showarticletitle{Direct multi-view multi-person 3d pose estimation}.
\newblock \bibinfo{journal}{\emph{Advances in Neural Information Processing Systems}}  \bibinfo{volume}{34} (\bibinfo{year}{2021}), \bibinfo{pages}{13153--13164}.
\newblock


\bibitem[Zhao et~al\mbox{.}(2024)]%
        {zhao2024single}
\bibfield{author}{\bibinfo{person}{Qitao Zhao}, \bibinfo{person}{Ce Zheng}, \bibinfo{person}{Mengyuan Liu}, {and} \bibinfo{person}{Chen Chen}.} \bibinfo{year}{2024}\natexlab{}.
\newblock \showarticletitle{A single 2d pose with context is worth hundreds for 3d human pose estimation}.
\newblock \bibinfo{journal}{\emph{Advances in Neural Information Processing Systems}}  \bibinfo{volume}{36} (\bibinfo{year}{2024}).
\newblock


\bibitem[Zhou et~al\mbox{.}(2019)]%
        {zhou2019continuity}
\bibfield{author}{\bibinfo{person}{Yi Zhou}, \bibinfo{person}{Connelly Barnes}, \bibinfo{person}{Jingwan Lu}, \bibinfo{person}{Jimei Yang}, {and} \bibinfo{person}{Hao Li}.} \bibinfo{year}{2019}\natexlab{}.
\newblock \showarticletitle{On the continuity of rotation representations in neural networks}. In \bibinfo{booktitle}{\emph{Proceedings of the IEEE/CVF conference on computer vision and pattern recognition}}. \bibinfo{pages}{5745--5753}.
\newblock


\bibitem[Zuo et~al\mbox{.}(2024)]%
        {Zuo_2024_CVPR}
\bibfield{author}{\bibinfo{person}{Chengxu Zuo}, \bibinfo{person}{Yiming Wang}, \bibinfo{person}{Lishuang Zhan}, \bibinfo{person}{Shihui Guo}, \bibinfo{person}{Xinyu Yi}, \bibinfo{person}{Feng Xu}, {and} \bibinfo{person}{Yipeng Qin}.} \bibinfo{year}{2024}\natexlab{}.
\newblock \showarticletitle{Loose Inertial Poser: Motion Capture with IMU-attached Loose-Wear Jacket}. In \bibinfo{booktitle}{\emph{Proceedings of the IEEE/CVF Conference on Computer Vision and Pattern Recognition (CVPR)}}. \bibinfo{pages}{2209--2219}.
\newblock


\end{thebibliography}

\appendix

\section{More Evaluations}

\subsection{Motion Tracking on Near-Template Shapes.} We also present quantitative results (Tab.~\ref{tab:totalcapture}) the TotalCapture \cite{trumble2017total} dataset. On action-rich real adult data, although the baseline method effectively performs motion tracking, our approach achieves higher accuracy, which we attribute to its shape-aware nature.
\subsection{Shape-error Analysis.} 
On three test datasets featuring children of varying heights, our method reduces joint position error by 23\%, 9\%, and 19\% compared to the second-best results, respectively (Tab.~\ref{tab:comparison_app}). The redirection of input shape-related data proves instrumental in this improvement. Additionally, for adults with larger body sizes (e.g., 191 cm compared to the 176 cm template), our approach consistently surpasses state-of-the-art performance across all evaluation metrics.

\begin{table}[ht]
    \centering
    \caption{Quantitative comparison results with state-of-the-art on TotalCapture.}
    \label{tab:totalcapture}
    \begin{minipage}{0.48\textwidth}
        \centering
        \begin{tabular}{c|ccccc}
            \toprule
            Method & SIP Err & Ang Err & Joint Err & Mesh Err & Jitter \\
            \midrule
            \multicolumn{6}{c}{TotalCapture} \\
            \midrule
            TransPose & 18.12& 14.91 & 7.10 & 8.09& 1.95\\
            PIP & 14.52 & 13.85 & 6.22 & 7.21 &\textbf{ 0.21} \\
            TIP & 15.62& 14.45 & 6.76 & 7.79 & 1.74\\
            ASIP & 13.45& 11.97& 5.28& 7.06& 0.23\\
            PNP & 11.36& 11.11& 4.89& 5.60& 0.32\\
            SAIP (ours) & \textbf{11.22}& \textbf{10.96}& \textbf{4.75}& \textbf{5.47}& 0.42\\
            \bottomrule
        \end{tabular}
    \end{minipage}%
    \hfill
\end{table}

\begin{table}[H]
    \centering
    \small 
    \caption{Quantitative comparison results with state-of-the-art methods on our shape-diverse subjects selected from the MID dataset.}
    \label{tab:comparison_app}
    \vspace{-5pt}
    \begin{tabular}{l|ccccc}
        \toprule
        Method & SIP Err & Angle Err & Joint Err & Mesh Err & Jitter Err \\
        \midrule
        \multicolumn{6}{c}{Subject 1 (height 118 cm)} \\
        \midrule
        TransPose & 28.26 & 12.41 & 8.68 & 10.25 & 0.05 \\
        PIP & 29.97 & 13.72 & 6.10 & 6.66 & 0.05 \\
        TIP & 31.24 & 22.48 & 6.83 & 7.68 & 0.09 \\
        ASIP & 36.56 & 14.26 & 9.59 & 10.31 & 0.10 \\
        PNP & 28.47 & 15.52 & 4.60 & 5.49 & 0.05 \\
        SAIP (ours) & \textbf{25.64} & \textbf{9.51} & \textbf{3.40} & \textbf{3.97} & \textbf{0.04} \\
        \midrule
        \multicolumn{6}{c}{Subject 2 (height 138 cm)} \\
        \midrule
        TransPose & 26.53 & 16.32 & 9.02 & 10.75 & 0.19 \\
        PIP & 24.62 & 13.32 & 10.13 & 11.03 & 0.10 \\
        TIP & 30.04 & 16.50 & 8.17 & 11.00 & 0.22 \\
        ASIP & 23.38 & 13.01 & 5.79 & 6.21 & 0.19 \\
        PNP & 24.40 & 13.25 & 8.07 & 7.11 & \textbf{0.09} \\
        SAIP (ours) & \textbf{20.58} & \textbf{8.44} & \textbf{5.08} & \textbf{5.81} & 0.10 \\
        \midrule
        \multicolumn{6}{c}{Subject 3 (height 144 cm)} \\
        \midrule
        TransPose & 26.44 & 14.19 & 11.52 & 13.84 & 0.13 \\
        PIP & 25.07 & 13.59 & 9.54 & 11.52 & 0.10 \\
        TIP & 37.98 & 15.67 & 12.48 & 13.10 & 0.19 \\
        ASIP & 26.14 & 12.68 & 10.52 & 11.82 & 0.11 \\
        PNP & 17.02 & 11.61 & 6.08 & 7.26 & \textbf{0.09} \\
        SAIP (ours) & \textbf{14.23} & \textbf{9.93} & \textbf{4.60} & \textbf{5.71} & \textbf{0.09} \\
        \midrule
        \multicolumn{6}{c}{Subject 4 (height 191 cm)} \\
        \midrule
        TransPose & 15.51 & 12.09 & 5.49 & 6.14 & 0.70 \\
        PIP & 14.92 & 10.27 & 4.23 & 5.02 & 0.12 \\
        TIP & 15.41 & 9.08 & 4.71 & 5.41 & 0.12 \\
        ASIP & 16.42 & 9.19 & 5.42 & 6.19 & \textbf{0.10} \\
        PNP & 15.38 & 10.62 & 4.61 & 5.40 & 0.12 \\
        SAIP (ours) & \textbf{13.12} & \textbf{7.40} & \textbf{4.18} & \textbf{4.82} & 0.12 \\
        \bottomrule
    \end{tabular}
\end{table}

\end{document}